\documentclass[useAMS,usenatbib]{mnras}

\usepackage[T1]{fontenc}
\usepackage{ae,aecompl}
\usepackage{array}

\usepackage{graphicx}
\graphicspath{{figures/}}	
\usepackage{amsmath}	
\usepackage{epstopdf}
\usepackage{multirow}
\usepackage{amssymb}

\usepackage{newtxtext,newtxmath}

\newcommand{\gaia}{\textit{Gaia}}	
\newcommand{\teff}{$T_{\rm{eff}}$}

\newcommand{\msolar}{M$_{\odot}$}

\title[Local WD stellar formation history]{Local stellar formation history from the 40\,pc white dwarf sample}

\author[Cukanovaite et al.]{E. Cukanovaite$^1$\,{\Huge \footnotemark},
P.-E. Tremblay$^1$, S. Toonen$^2$, K.D. Temmink$^3$, Christopher J. Manser$^4$,
\newauthor{M.W. O'Brien$^1$ and J. McCleery$^1$}
  \\
$^{1}$ Department of Physics, University of Warwick, Coventry CV4 7AL, UK \\ 
$^{2}$ Anton Pannekoek Institute for Astronomy, University of Amsterdam, 1090 GE Amsterdam, The Netherlands \\
$^{3}$ Department of Astrophysics/IMAPP, Radboud University Nijmegen, P.O. Box 9010, 6500 GL Nijmegen, The Netherlands \\
$^{4}$ Astrophysics Group, Department of Physics, Imperial College London, Prince Consort Rd, London, SW7 2AZ, UK
}

\date{Accepted XXX. Received YYY; in original form ZZZ}

\pubyear{2022}

\begin{document}
\label{firstpage}
\pagerange{\pageref{firstpage}--\pageref{lastpage}}
\maketitle

\begin{abstract}
We derive the local stellar formation history from the \gaia-defined 40\,pc white dwarf sample. This is currently the largest volume-complete sample of white dwarfs for which spectroscopy is available, allowing for classification of the chemical abundances at the photosphere, and subsequently accurate determination of the atmospheric parameters. We create a population synthesis model and show that a uniform stellar formation history for the last $\approx10.5$\,Gyr provides a satisfactory fit to the observed distribution of absolute \gaia~$G$ magnitudes. To test the robustness of our derivation, we vary various assumptions in the population synthesis model, including the initial mass function, initial-to-final mass relation, kinematic evolution, binary fraction and white dwarf cooling timescales. From these tests, we conclude that the assumptions in our model have an insignificant effect on the derived relative stellar formation rate as a function of look-back time. However, the onset of stellar formation (age of Galactic disc) is sensitive to a variety of input parameters including the white dwarf cooling models. Our derived stellar formation history gives a much better fit to the absolute \gaia~$G$ magnitudes than most previous studies.  
\end{abstract}


\footnotetext{E-mail: E.Cukanovaite.1@warwick.ac.uk}



\section{Introduction}

Clues of the past and future of the universe can be found by studying galaxies at different redshifts and therefore at different epochs of time (see, for example, \citealt{speagle2014,somerville2015,frebel2015,kruijssen2019} and references therein). Our own Milky Way is a perfect representative of spiral galaxies in the low-redshift universe (see, for example, \citealt{helmi2008,hou2014,helmi2020} and references therein) offering the additional advantage of being the only galaxy that can be studied from within. By examining the evolution of the individual stars in the Milky Way, we can learn how the baryonic matter of galaxies has developed over time.

The overall structure of the Milky Way is thought to be made up of four main baryonic components: the bulge and the bar, the thin disc, the thick disc, and the halo \citep{helmi2020}. It is agreed that the Milky Way follows the $\Lambda$CDM cosmological model \citep{white1978,davis1985,delucia2006,planck2016,bullock2017}, where galaxies form inside dark matter halos, attracting baryonic matter over time. The baryonic matter then cools and collapses towards the centre of the dark matter halo, forming a disc due to conservation of angular momentum \citep{mo1998,cole2000,keres2005}. Most importantly, the galaxy also grows via mergers with neighbouring galaxies \citep{odenkirchen2001,ibata2002,niederste2010,besla2010,belokurov2013,grillmair2016,helmi2020}, with the first point of contact being the outer halo of the Milky Way. Nevertheless, the mergers have an impact on all the components of the Galaxy. These mergers leave imprints in the form of Galactic streams, which have been observed and studied extensively \citep{belokurov2013,grillmair2016,helmi2020}. 

It is accepted that all the components of the Galaxy are interlinked in their evolution, due to a single Galactic potential \citep{guedes2013} and radial migration of stars as they get older \citep{sellwood2002,siebert2008,bond2010,minchev2010,minchev2015}. Thus, some studies have trouble separating the different components \citep{norris1991,bovy2012a,bovy2012b,hayden2017}. It is now accepted that kinematically the thin and thick disc stars are hard to differentiate, yet, the two components can be separated by stellar metallicity \citep{minchev2015,kawata2016,helmi2020}, with the members of the thick disc being more metal-poor and therefore older \citep{gilmore1989,haywood2013,hayden2015}. Similar issues have recently been uncovered in trying to separate the thick disc from the inner halo, as both are thought to have been significantly affected by the Gaia-Enceladus merger, and are observed to overlap in the Solar neighbourhood \citep{koppelman2018,belokurov2018,haywood2018,helmi2018,gallart2019,myeong2019,helmi2020}. However, it can be summarised that the oldest stars are found in the halo, which was formed around 10-12 Gyr ago \citep{salaris2002,jofre2011,kilic2017,gallart2019}. On the other hand, the thin disc is the youngest, formed around 8 Gyr ago \citep{jimenez1998,liu2000,delpeloso2005a,delpeloso2005c,haywood2013,kilic2017,tononi2019}. Ultimately, it is apparent that the evolution of the Milky Way is complex and is not yet fully understood, but has recently been and continues to be aided by the data coming from the \gaia~satellite \citep{gaia2016,gaia2016b,gaia2018,gaia2021}. 

The evolution of the Milky Way can be summarised in a stellar formation history, which denotes how many stars were formed during a particular period. Traditionally, its derivation has relied on methods involving main sequence stars, but other types of stars can also be used. Some example methods include nucleocosmochronometry, isochrone fitting, and asteroseismology (see, for example, the review of \citealt{soderblom2010} and references therein). An independent and alternative method is based on white dwarf stars (see, for example, \citealt{winget1987,rowell2013,tremblay2014,kilic2017,isern2019,fantin2019}). These stars do not undergo any significant nuclear fusion in their cores and instead cool with age via the gravothermal process, providing accurate age estimates \citep{althaus2010,rebassa2015,heintz2022}. Additionally, white dwarfs encompass information over the entirety of the Milky Way’s history, with the oldest known white dwarfs having cooling ages of around 10\,Gyr, meaning that they have come from main sequence progenitors that are even older \citep{fontaine2001}.

The first established method of using white dwarfs to determine the stellar formation history involved their luminosity function \citep{winget1987}, where a sharp drop-off is observed at dim magnitudes. This was understood as the finite age of the Milky Way and it being smaller than the time needed for a white dwarf to cool down to magnitudes dimmer than the drop-off in the luminosity function. This method has been applied both to the Galaxy as a whole \citep{winget1987,garciaberro1988,bessell1993} and to the individual Galactic components (see, for example, \citealt{mochkovitch1990,isern1998,garciaberro2010,kilic2017,torres2019,fantin2019,fantin2021,torres2021} and reference therein). Furthermore, these and other studies have also shown that the bright end of the white dwarf luminosity function is also highly sensitive to the stellar formation history \citep{noh1990,garciaberro2016}. Thus, the white dwarf luminosity function tracks the entirety of the Galactic stellar formation history and can be manipulated to get more accurate results. For example, \cite{isern2019} and references therein showed that massive white dwarfs can be used to find stellar formation histories without needing to invoke uncertainties from the main sequence lifetimes.

For either white dwarf or main sequence methods, the derivation of an accurate stellar formation history is heavily impeded by various observational biases. In particular, to overcome incompleteness of observational samples, population synthesis codes have to include complex bias corrections, which, for example, can be functions of survey pointing, magnitude and colour. With the advent of \gaia~and its data releases, the completeness of the night sky has increased significantly \citep{gaia2016,gaia2016b,gaia2018,gaia2021} with unprecedented precision in astrometry \citep{lindergren2021a,lindergren2021b}. One such tremendous improvement involved the careful cataloguing of white dwarfs in \gaia~Data Release 2 (DR2, \citealt{gentilefusillo2019}) and early Data Release 3 (eDR3, \citealt{gentilefusillo2021}). 

In this paper we aim to derive the local stellar formation history using a subset of 1083 \gaia~white dwarfs within 40\,pc of the Sun. This subset is estimated to have a \textit{Gaia} completeness of around 96\% \citep{tremblay2020,mccleery2020}. Furthermore, \cite{mccleery2020}, \cite{tremblay2020} and \cite{obrien2022} have compiled the atmospheric compositions for nearly all white dwarfs ($>$96\%) in that sub-sample. They used \gaia-independent spectroscopic observations, making the 40\,pc sample the largest, most complete volume of spectroscopically confirmed white dwarfs to date. This is particularly important as the atmospheric composition of a white dwarf directly affects its temperature, radius and mass derived from the \gaia~data \citep{bergeron2019}, leading to errors of up to 20\% in these parameters if the atmospheric composition is instead assumed. Another advantage of this sample compared to the larger volume samples is the negligible amount of interstellar dust, allowing for a direct comparison of observed and predicted \gaia~$G$ magnitudes. Additionally, for larger samples the \gaia~errors become larger, making the parameters provided by \gaia~less accurate, as well as making the sampled volume harder to define.

In Sect.~\ref{sec:40pc} we introduce the 40\,pc \gaia~sample of white dwarfs. Sect.~\ref{sec:numerical} describes the simulation constructed to model the local white dwarf sample. Sect.~\ref{sec:fitting} describes the method used to derive the local stellar formation history. We compare our results to other studies in Sect.~\ref{sec:other_sfh}. In Sect.~\ref{sec:ass_bias} we discuss the various uncertainties associated with assumptions made in the simulation. We discuss the results and conclude in Sect.~\ref{sec:disc}.
 
\section{The 40\,pc sample of white dwarfs}~\label{sec:40pc}

To derive the local stellar formation history, we use the northern 40\,pc sample of spectroscopically confirmed white dwarfs curated by \cite{mccleery2020} and the southern 40\,pc sample curated by \cite{obrien2022}. The northern catalogue was originally published based on DR2, but has recently been updated with \gaia~eDR3 (\citealt{gentilefusillo2021})\footnote{\url{https://cygnus.astro.warwick.ac.uk/phrgwr/40pcTables/index.html}}. In total, the 40\,pc sample contains 1083 white dwarfs: 626 have hydrogen-dominated atmospheres; 266 have helium-dominated atmospheres; 9 are likely helium-rich DC white dwarfs with strong CIA absorption. The remaining 182 white dwarfs have spectra that show no significant absorption lines and have \gaia\ photometric \teff\ below $\approx$ 5000\,K, where both He and H lines become invisible. Therefore, these cool white dwarfs have unconstrained atmospheric compositions (see \citealt{mccleery2020} for more detail).

Given the precision of the \gaia~data, the observed absolute $G$ magnitude provides a robust independent parameter for the determination of the local stellar formation history. The absolute magnitude is calculated from the distance modulus equation, where the distance is assumed to be equal to the inverse of parallax. This is not a correct assumption for parallaxes with large errors because of the data processing done by \gaia~\citep{bailerjones2018,bailerjones2021}. However, as our sample is local and has small parallax errors, this assumption is valid, and as such we utilise it. The distance modulus equation also relies on the observed apparent $G$ magnitude, which given the locality of our sample, we do not deredden, as dust becomes only important for distances larger than around 50\,pc \citep{gentilefusillo2021}. 

Our approach is similar to earlier studies deriving the local stellar formation history from the white dwarf luminosity function, but does not rely on the determination of individual atmospheric parameters, the effective temperature ($T_{\rm eff}$) and surface gravity, which were necessary in the pre-\gaia~era to infer the distance, and hence luminosity. In particular, our method only depends on the absolute optical \gaia~$G$ flux and does not use \gaia~colours (or \gaia~inferred $T_{\rm eff}$), which are much more sensitive to the details of the modelled atmosphere, and thus the opacities. Nevertheless, any significant offset between observed and predicted \gaia~colours, such as that seen for white dwarfs with $T_{\rm eff} \lesssim 5000$\,K which suffer from a low-mass problem \citep{hollands2018,mccleery2020}, could have a smaller repercussion on the absolute \gaia~$G$ flux.

\begin{figure}
	\includegraphics[width=\columnwidth]{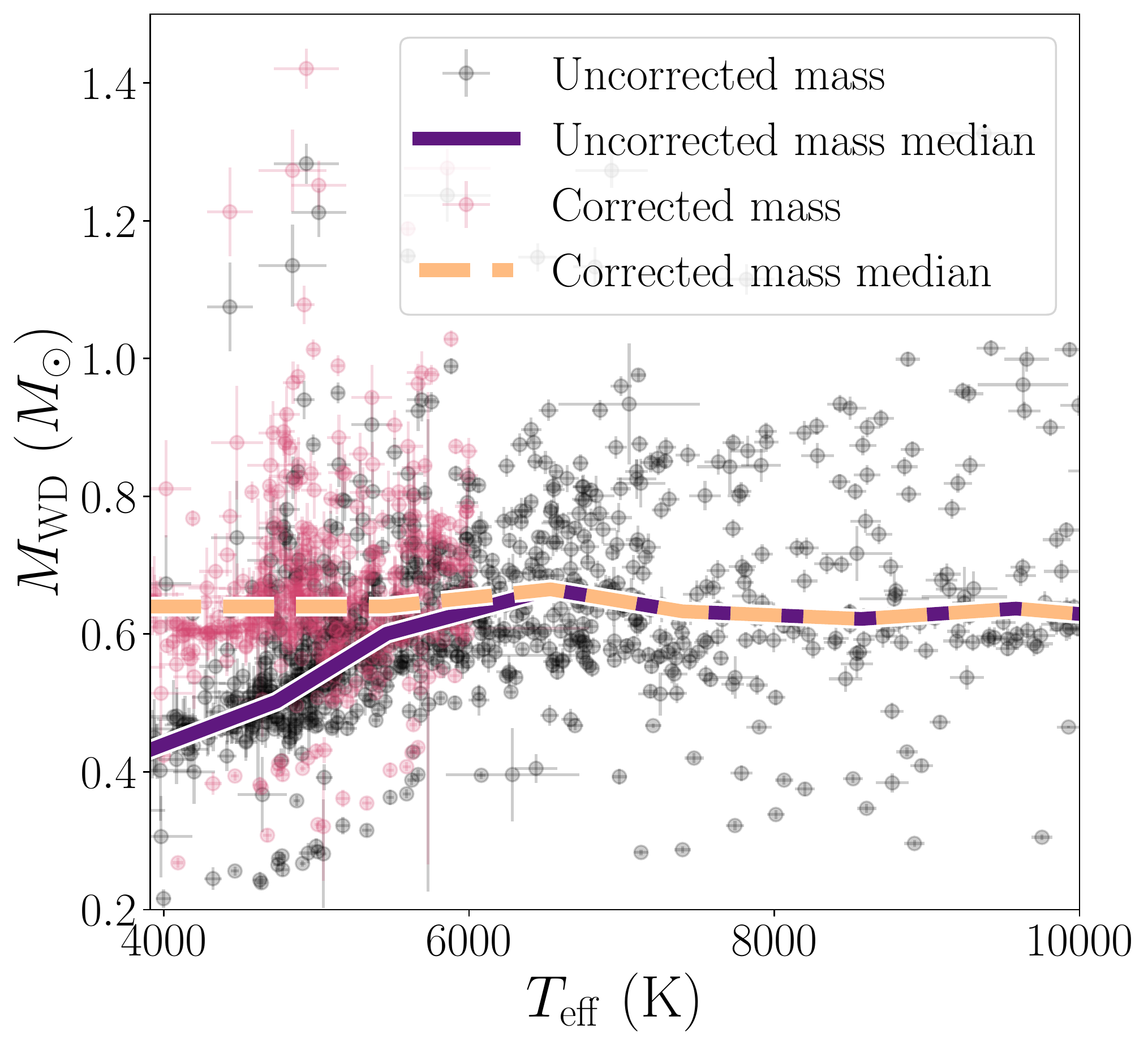}
    \caption{The mass-$T_{\rm eff}$ distribution of the all-sky 40\,pc white dwarf sample. The black dots represent the original masses of the white dwarfs, with the solid purple line representing the median mass in bins of around 1000 K. Due to the unrealistic downturn in the median mass below 6000 K, the masses had to be corrected. The corrected masses are denoted by blood orange dots (see Table\,\ref{tab:mass_corr}), with the corrected median mass indicated by the dashed orange line.} 
    \label{fig:masses}
\end{figure}

In Fig.~\ref{fig:masses} the mass-$T_{\rm eff}$ distribution of the 40\,pc sample is shown as black dots. These parameters are derived using the method described in \cite{mccleery2020}, which is based on eDR3 colours and the appropriate atmospheric composition given the identified spectral type. 

There is a lower limit to the white dwarf mass for single star evolution, since the creation of a white dwarf with a mass smaller than around 0.54\,\msolar~would take longer than the current age of the Universe (see, for example, \citealt{cummings2016}). Thus, the very low observed masses in Fig.~\ref{fig:masses} are caused by binary systems. In the case of unresolved double white dwarf systems, the observed \gaia~flux will appear over-luminous when fitting with a single white dwarf model, resulting in a large radius and low white dwarf mass \citep{mccleery2020}. Another scenario is when a companion strips the mass of a white dwarf during a common-envelope evolution, leaving a white dwarf with a mass that is smaller than what is possible assuming single star evolution (see, for example, \citealt{temmink2020}). A third path for binary evolution is a merger, resulting in an apparently single white dwarf of any mass \citep{temmink2020}. We will address white dwarf mergers in a later section when describing our simulation.

Unrelated to the above point, from Fig.~\ref{fig:masses} we can clearly see that there is a downturn to lower masses for $T_{\rm eff} \lesssim 6000$\,K or a cooling age of about 2.5\,Gyr. It is better illustrated by the purple solid line in Fig.~\ref{fig:masses}, which represents the median mass in a given \teff~bin. It is not consistent with evolutionary models of white dwarfs, which predict cooling at almost constant total mass and radius. In particular, Galactic population simulations from \citet{tremblay2016} have shown that the mean mass of white dwarfs is predicted to stay the same within 1\% for almost all look-back times, except for the first $\approx$ 500\,Myr after the Galactic disc formation, where the mean mass of white dwarfs was slightly higher than it is now because they came on average from slightly more massive progenitor stars.   

The explanation for the small masses of cool white dwarfs is likely attributed to an opacity which is missing or is incorrect in the atmospheric models \citep{bergeron2019}. In the absence of appropriate grids of model atmospheres, we take the following ad-hoc approach. For each bin below 6000\,K, we find the difference between the median mass in the bin, and the overall median mass above 6000\,K, but below 20\,000\,K. We then apply this difference to the masses of all white dwarfs below 6000\,K as indicated by blood-orange points in Fig.~\ref{fig:masses}. This means that the median mass below 6000\,K is then corrected to agree with evolutionary models. This corrected median mass is denoted in Fig.~\ref{fig:masses} by the dashed orange line. The specific equation for the mass correction can be found in Tab.~\ref{tab:mass_corr}.

Now that white dwarf masses are corrected for the missing opacity, we can remove any white dwarf whose corrected mass is still below the single star evolution lower mass limit at 0.54\,\msolar. This removes 110 white dwarfs, leaving us with a sample of 973 white dwarfs. 

\begin{table}
	\centering
	\caption{The mass correction applied to correct for issues with opacity. The table shows the amount of mass that needs to be added to individual white dwarfs for specific \teff~ranges, in order to get the median mass to agree to the median masses above 6000~K, but below  20\,000~K, as expected from evolutionary models.}
	\label{tab:mass_corr}
	\begin{tabular}{|l|r|} 
  \hline
  Correction in mass & \teff~range correction\\ 
  (\msolar) & applies to (K) \\
  \hline
  $+$0.208 & $3000\le$~\teff~$<4000$ \\
  
  $+$0.138 & $4000\le$~\teff~$<5000$ \\
  
  $+$0.0395 & $5000\le$~\teff~$<6000$ \\
  \hline
	\end{tabular}
\end{table}

\begin{figure}
	\includegraphics[width=\columnwidth]{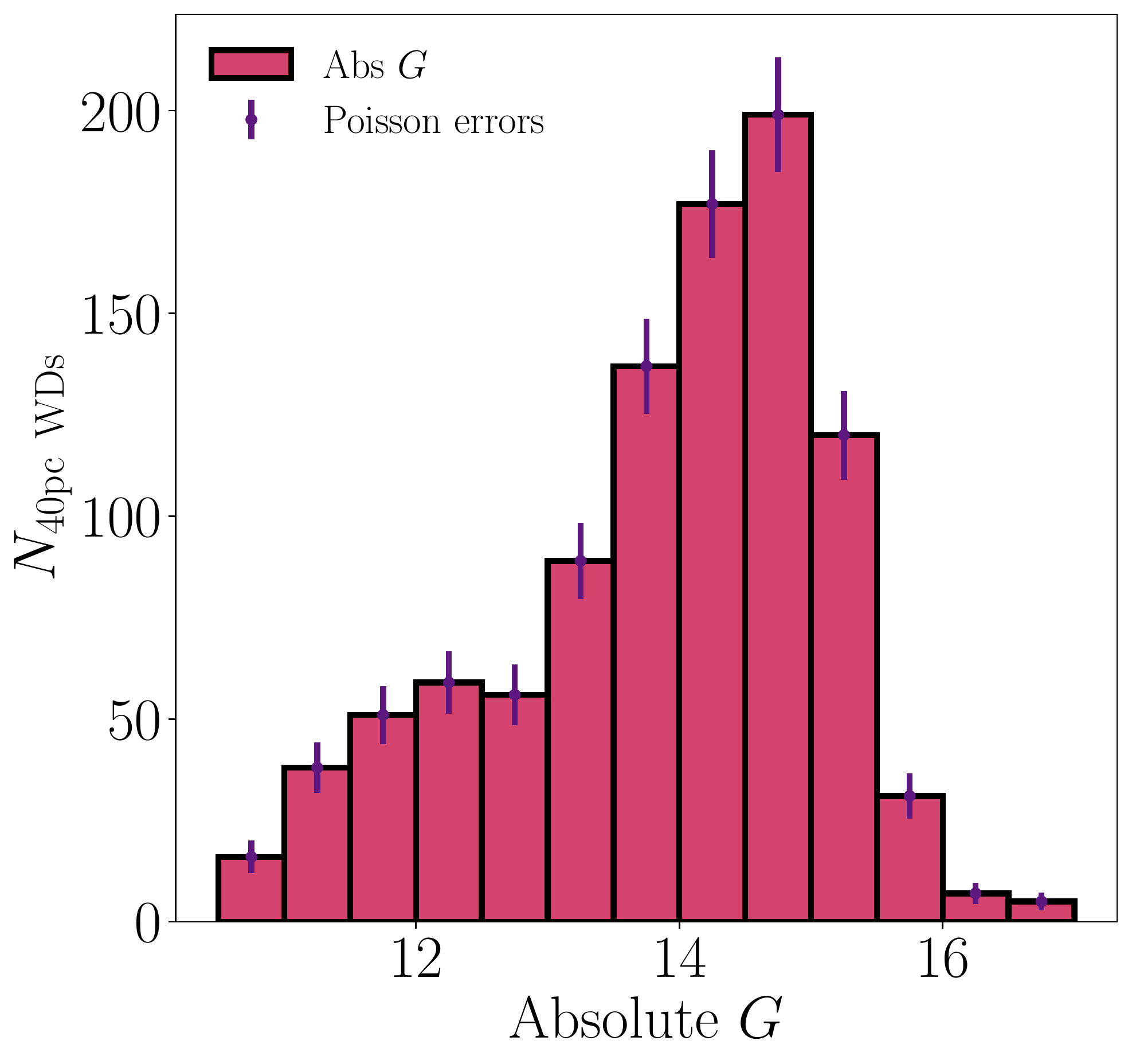}
    \caption{A histogram of the absolute $G$ magnitudes for the 40\,pc white dwarf sample. The histogram is plotted in solid pink. Given the precision of \gaia~data, the individual errors on absolute $G$ magnitudes are insignificant. Thus, we plot the Poisson errors in purple to account for the dominant source of errors. The sizes of the bins are 0.5\,dex in $G$ magnitude.} 
    \label{fig:absG_hist}
\end{figure}

In Fig.~\ref{fig:absG_hist} we show a histogram of the absolute magnitude data we will be using to determine the local stellar formation history. This is essentially the white dwarf optical wavelength-range luminosity function for the 40\,pc sample. Given the precision of \gaia, the errors on absolute $G$ magnitude are insignificant, with a median value of 0.03\%. Instead, we plot the Poisson errors. As expected, we see a sharp drop-off at dimmest magnitudes, which is due to the finite age of the Galaxy. The histogram seems to have two peaks, one centred around $G=12$ and the other around $G=14.5$. This has been seen in other white dwarf luminosity functions (see, for example, \citealt{harris2006}) and is due to the different cooling mechanisms white dwarfs experience as they age \citep{koester1990}. 

\subsection{Separating the different Galactic populations}

To derive the stellar formation history, we must first have some idea of the shape it might take. As there are three Galactic components that can exist in our local volume, the question arises whether they each have their own separate history. To answer this, we must first see if the three components are apparent in our sample. There are several methods used in literature to identify and separate the populations of the thin disc, the thick disc, and the halo of the Galaxy. They rely on the assumption that the components were formed at different times, but ages can be difficult to measure for main-sequence stars. Hence, the proxies of age, such as kinematics and/or metallicity can be used to differentiate the populations, since, for example, older stars are formed in lower metallicity environments and have larger velocity dispersions than younger stars \citep{aumer2009,casagrande2011}. When relying on observations of white dwarfs, the metallicity information of the main sequence star is lost due to the high gravitational fields of white dwarfs, which cause all but the lightest elements to sink out of the atmosphere towards the core. Therefore, in the following we aim to use the kinematics of white dwarfs to assign them to a specific Galactic population. For the halo population, the typical Galactic velocities in the local standard of rest are around 180-220 km s$^{-1}$ (see, for example, \citealt{majewski1992,smith2009,nissen2010,du2018} and references therein). The velocities of the thick disc stars are expected to be smaller than the halo velocity, but larger than 70 km s$^{-1}$ \citep{bensby2014}. Thin disc stars are considered to have velocities below 70 km s$^{-1}$. Thus, we will be using these values in the following analysis. In particular, we will use 180 km s$^{-1}$ for the halo population in order to determine the maximum possible number of halo white dwarfs.

From \gaia~eDR3 we have access to the proper motions, right ascensions, declinations, and distances for the white dwarfs in the 40\,pc sample. If we also have access to their radial velocities, we can derive their Galactic velocity, where the component $U$ is directed towards the Galactic center, component $V$ is in the direction of Galactic rotation, and component $W$ is perpendicular to the Galactic disc.  These components are corrected for the local standard of rest. The first step is therefore to compile all available radial velocities for our sample. To achieve this we used the radial velocity catalogues of \cite{anguiano2017} and \cite{napiwotzki2020}\footnote{Note that we also looked at samples of \cite{pauli2006} and \cite{raddi2021}, but in the former case, the radial velocities were older measurements and had bigger errors than the ones found from \cite{anguiano2017} and \cite{napiwotzki2020}, and in the latter case there were no radial velocities for our white dwarfs.}, which have been corrected for gravitational redshifts of white dwarfs. In total, we found 64 (5.9\% of total sample) northern and 59 (5.4\% of total sample) southern white dwarfs with radial velocities, all with hydrogen-dominated atmospheres. Therefore, 11.4\% of the white dwarfs in the 40\,pc sample had available radial velocities. If both \cite{anguiano2017} and \cite{napiwotzki2020} samples had available radial velocities for a given white dwarf, we chose the \cite{napiwotzki2020} data, due to the sample's smaller errors and recentness. With the available radial velocities, we calculate the Galactic velocities by utilising Astropy and its ICRS, SkyCoord, GalacticLSR and Units packages and classes \citep{astropy2013,astropy2018}. For white dwarfs with no available radial velocities, we instead randomly sample a velocity from the radial velocity distribution of \cite{raddi2021}. Once a radial velocity is assigned, we follow the same procedure as before to find the Galactic velocity components. However, to reduce any impact of randomness, we sample 1000 radial velocities for each white dwarf, and use bootstrapping to determine the median values of the Galactic velocity components and their associated errors. The errors are determined at the confidence limits of 2.5\% and 97.5\%.

\begin{figure}
	\includegraphics[width=\columnwidth]{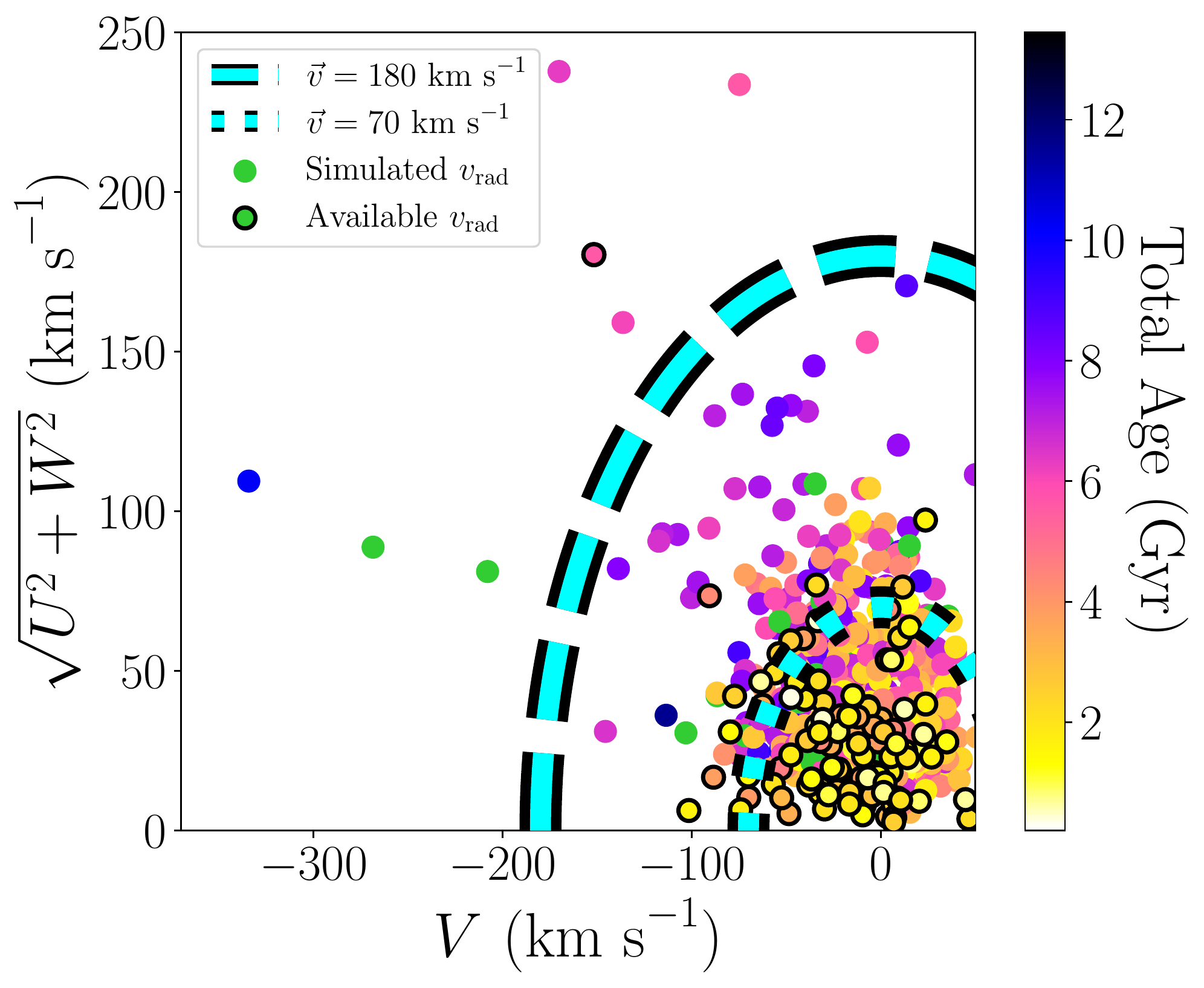}
    \caption{The Toomre diagram for the 40\,pc sample of white dwarfs. The plotted velocity components are Galactic and have been calculated by taking into account the local standard of rest. The dashed and dotted cyan lines denote the total velocities equal to 180 km s$^{-1}$ and 70 km s$^{-1}$, which are reference values that can be used to identify the halo and thick disc populations, respectively. The colour of the individual points illustrates the derived total age of a given white dwarf, as indicated by the right colour bar. The green markers symbolize white dwarfs whose ages were larger than the age of the Galaxy. Black contours represent white dwarfs with radial velocities found from \protect\cite{anguiano2017} and \protect\cite{napiwotzki2020}, while objects with no contours have simulated radial velocity from the sample of  \protect\cite{raddi2021}.}
    \label{fig:toomre}
\end{figure}

Fig.~\ref{fig:toomre} shows the Toomre diagram for our 40\,pc sample. The lower limits of the total velocity for the thick disc and the halo populations are plotted as dotted and dashed cyan lines, respectively. The \gaia\ errors are smaller than the individual data points and the main source of error comes from the observed radial velocity data, with the median error being around 8\%. In terms of the simulated radial velocities, the distribution of \cite{raddi2021} is well peaked, leading to a small error on the median Galactic velocities. To test the effect of the more uncertain radial velocities, we also computed Toomre diagrams where all the radial velocities were set to 0 km s$^{-1}$ and where the \cite{raddi2021} distribution is shifted by several tens of km s$^{-1}$. We find no significant difference in the overall appearance of the plot, unless the shift is of the order of 50 km s$^{-1}$, which is improbable as it would make the simulated velocities much higher than the velocities for the white dwarfs with observed radial velocities.

The colours of the markers in Fig.~\ref{fig:toomre} indicate the total ages of the white dwarfs. The white dwarf ages were derived using the University of Montr\'eal cooling tables \citep{bedard2020}\footnote{\url{http://www.astro.umontreal.ca/~bergeron/CoolingModels}}. In the 40\,pc sample, each white dwarf has a determined atmospheric composition from observed spectra, therefore the appropriate cooling table (thick or thin hydrogen layer) was used to determine the cooling age. Each white dwarf also has a mass derived from \gaia~data (corrected for opacity issues in the previous section), which is transformed into the mass of its progenitor using the initial-to-final mass relation of \citealt{elbadry2018}. This progenitor mass allows for the determination of the main sequence lifetime using the \cite{hurley2000} relation at solar metallicity, which is then combined with the cooling age to find the total age. Some white dwarfs are denoted by green colours as their ages could not be determined. This is because either the atmospheric parameters were not available from the samples of \cite{mccleery2020} and \cite{tremblay2020}; or their total age was larger than the age of the universe. In total 78 white dwarfs do not have age determinations.

Overall, the Toomre diagram shows that all three potential regions of Galactic components are populated, with the regions of the thin and the thick disc components being most populous. This is expected because the thin disc is concentrated in the Galactic plane where the Sun is found, while the thick disc and the halo components are more spread out. The figure also indicates that white dwarfs with lower velocities are more likely to be young. This is expected because of the age-velocity dispersion relation. We must note that by using the Toomre diagram we can identify potential candidates of the different Galactic components, but we cannot separate them with absolute certainty. The diagram seems to indicate there are two distinct velocity populations in our sample: one concentrated ($\lesssim$100 km s$^{-1}$) and one diffuse component ($\lesssim$300 km s$^{-1}$) in the velocity space, with the diffuse component mostly consisting of old stars (total age $\gtrsim$ 6 Gyr). From the data available, it is difficult to assign these components to specific Galactic populations such as thin and thick discs or halo, and these populations may overlap in velocity space. As mentioned in the Introduction, other studies have also found an overlap between the thin and thick discs in the local neighbourhood, possibly due to the Gaia-Enceladus merger. Additionally, the white dwarf study of \cite{torres2019}, which used \gaia~data, has found 21 halo white dwarfs within 40\,pc, suggesting the presence of the three Galactic components in this volume. That is more halo white dwarfs than we find, however, their classification technique is more advanced than the technique used in this paper. In some cases they classify white dwarfs with velocities smaller than 180 km s$^{-1}$ as halo white dwarfs, based on their other intrinsic parameters. We therefore proceed by testing stellar formation histories with one, two and three Galactic components to account for all possible scenarios.

\subsection{Age-velocity dispersion}~\label{sec:vel_bias}

The age-velocity dispersion relation introduces an observational bias which we must consider when comparing Galactic simulations with the observations. Older white dwarfs are statistically more likely to have larger velocities and as such are less likely to be observed in a given volume close to the Galactic plane. If this bias was not taken into account, it could lead to an erroneous conclusion that the stellar formation history must have peaked at recent times to explain an apparent over-abundance of younger white dwarfs. In this work we consider this bias to affect the $W$ component of the total velocity only, as we assume to first order that the same amount of objects come in to and out of 40\,pc volume in the directions of the $U$ and $V$ velocity components due to radial mixing.

\begin{figure}
	\includegraphics[width=\columnwidth]{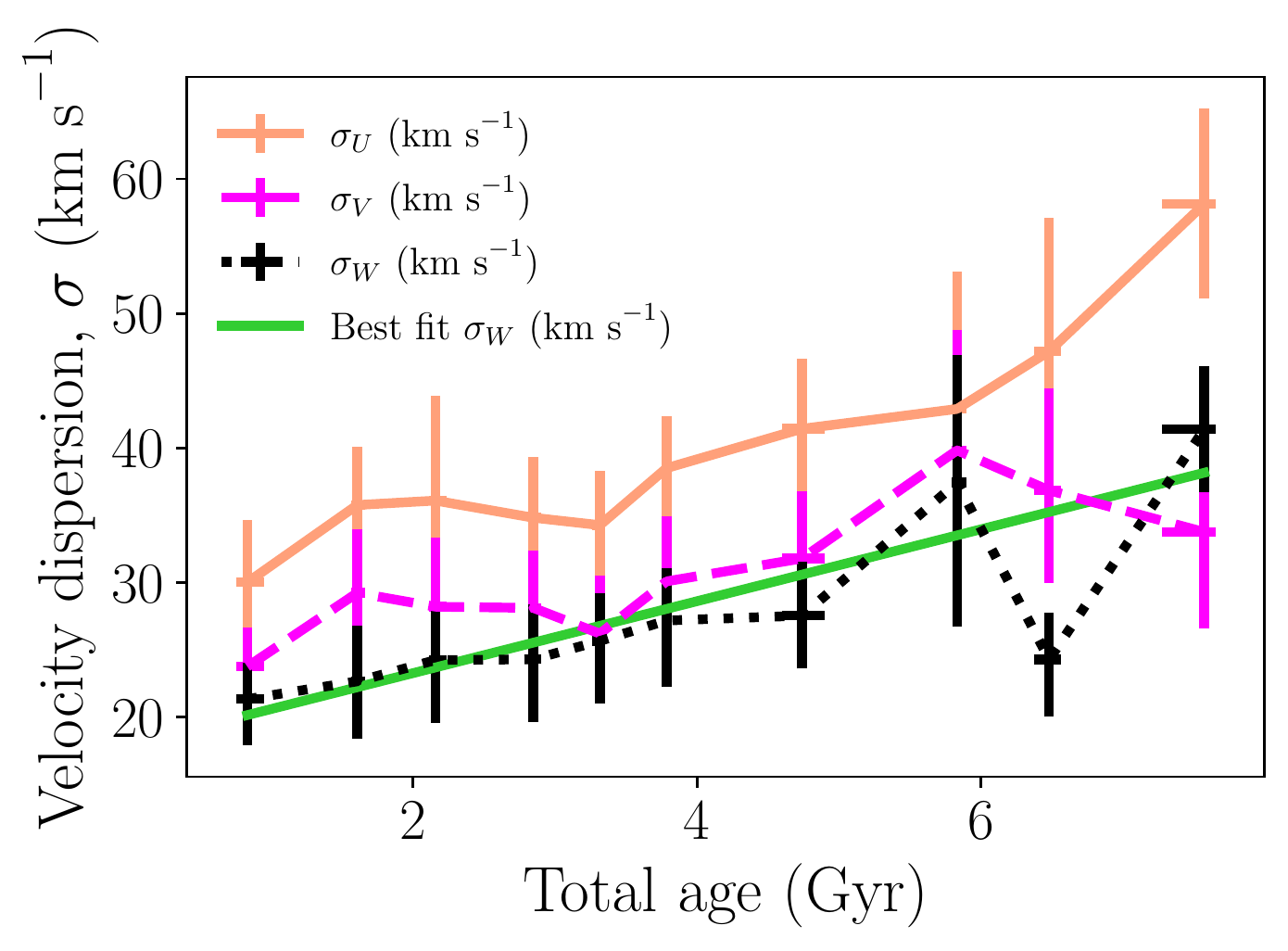}
    \caption{The binned age-velocity dispersions for the 40\,pc sample of white dwarfs. These relations are plotted for the three Galactic velocity components, $U$, $V$ and $W$, in solid orange, dashed pink, and dotted black lines, respectively. Median total ages and errors are found for each bin using bootstrapping. The velocity dispersions are also found using bootstrapping, taking into account that for most white dwarfs the radial velocities were simulated using the \protect\cite{raddi2021} radial velocity distribution. The errors shown on the plot represent the confidence intervals of 2.5\% and 97.5\%.}
    \label{fig:velocity_dispersion_vs_age_observed}
\end{figure}

In Fig.~\ref{fig:velocity_dispersion_vs_age_observed} we plot binned values of velocity dispersion and total age for the 40\,pc white dwarf sample. We use ten bins with nearly equal number of white dwarfs. We remove white dwarfs whose ages could not be determined. For each bin we find the median total age and its error using bootstrapping, where the value of the error is taken as the confidence intervals of 95.7\% and 2.5\%. To determine the velocity dispersion, we also use bootstrapping with the same confidence intervals as errors. Bootstrapping is beneficial in our case, as the errors on individual measurements are almost insignificantly small when compared to the actual scatter of the data points. Bootstrapping also allows us to easily account for the random fluctuations arising from using simulated radial velocities.

In line with other studies \citep{tremblay2014,tremblay2016,mccleery2020,raddi2021}, Fig.~\ref{fig:velocity_dispersion_vs_age_observed} shows the velocity dispersion increasing with age. The range of the velocity dispersion, 10-60 km s$^{-1}$ is also consistent with the \cite{raddi2021} sample. Additionally, we find that the $U$ component has the largest velocity dispersion. 
Overall, we find that the velocities of the 40 pc white dwarf sample resemble previous studies of white dwarfs, and the main sequence stars, but further study is needed, since not all white dwarfs in our sample have observed radial velocities. We include the age-velocity dispersion in our simulation, as described in the next section, Sect.~\ref{sec:numerical}.

\section{Numerical simulation of the 40 pc white dwarf sample}~\label{sec:numerical}

The simulation of the local white dwarf sample is based on Monte Carlo methods. To ensure that the random number fluctuations are insignificant, we must generate a large enough number of simulated white dwarfs, which we fix at 3000 white dwarfs. We find that the fluctuations between the different simulations generated with this number of white dwarfs are on 2\% level. 
We define the total age of the star as the time difference from the present time and the time of stellar formation (formation time). The time variable used throughout this work is the lookback time, going back from present day ($t = 0$) to formation time. In the simulation, the formation time is assigned based on the assumed stellar formation history. To generalise the stellar formation history in a way that is suitable for fitting observed data, we assume either one, two or three Galactic components that each have their own uniform stellar formation history. The specific details on this can be found in Sect.~\ref{sec:fitting}.

For each white dwarf, a random position is generated in the three Cartesian ($x,y,z$) coordinates. Both the $x$- and $y$-coordinates are sampled from a uniform distribution which spans the distances between 0 and 40\,pc from the Sun. As mentioned in Sect.~\ref{sec:vel_bias}, we assume to first order that the amount of stars coming in and out of the 40\,pc in $x$ and $y$ directions are equal. On the other hand, the $z$-coordinate is sampled from a quasi-exponential distribution, which has been offset by 20\,pc, due to the Sun's vertical position relative to the Galactic centre (see, for example, \citealt{siegert2019}), i.e. from the probability density
\begin{equation}
\rho{(z)} = 1-\exp{(-\mathrm{(z+20})/\mathrm{scale \ height)}}.
\end{equation}

\begin{figure}
	\includegraphics[width=\columnwidth]{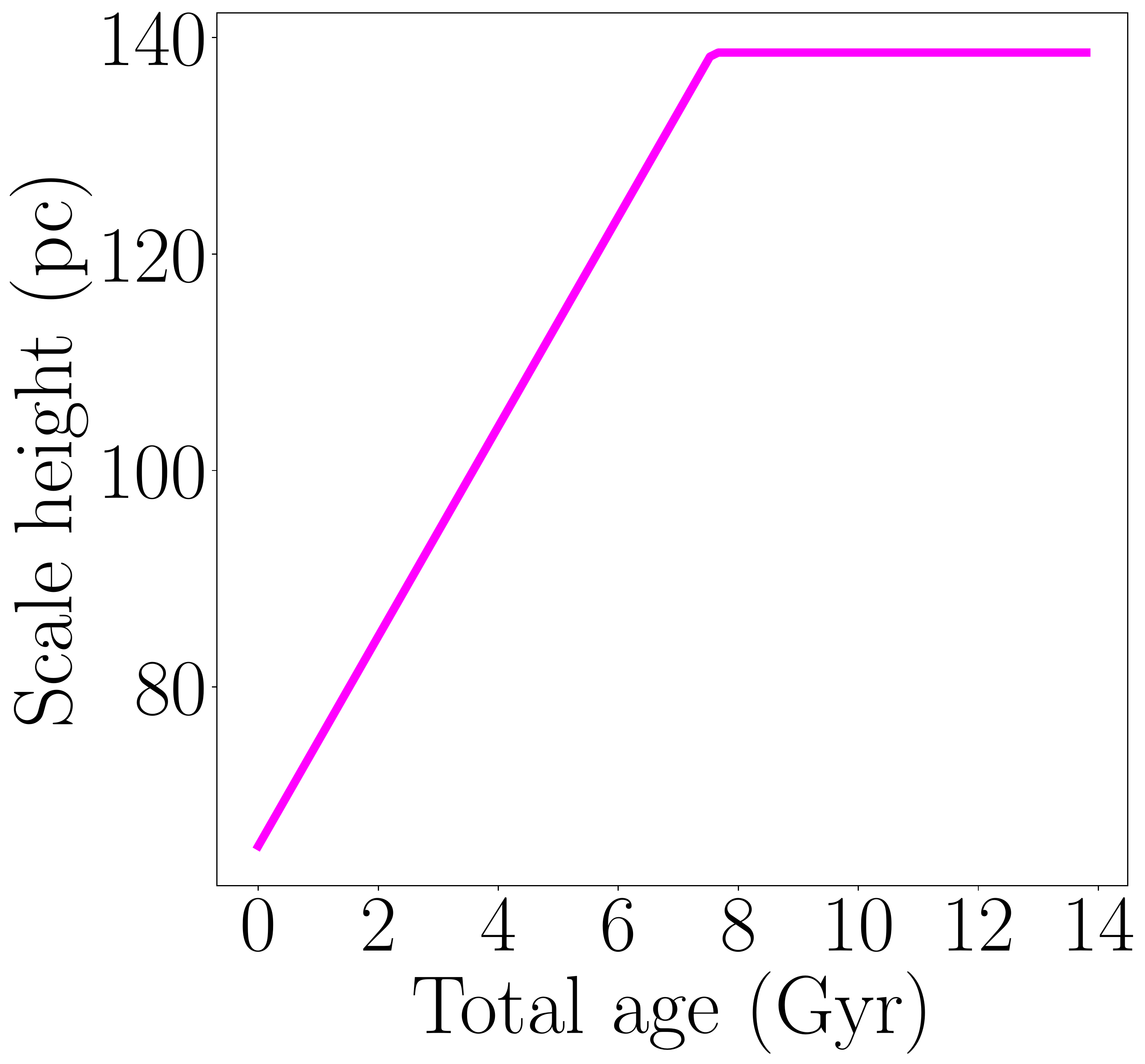}
    \caption{The scale height as a function of total age of the star. This relation has been derived based on a smooth monotonic fit of observed velocity dispersion of the $W$-velocity component, and the assumption that at age 1\,Gyr the stars have vertical scale height of 75\,pc \protect\citep{wegg2012,buckner2014}.}
    \label{fig:scale_height}
\end{figure}

The $z$-distribution is limited to distances between 0 and 40 pc. An important assumption in this distribution is the scale height of the Galactic component. In Sect.~\ref{sec:vel_bias} we found the relation between the velocity dispersion of the $W$ velocity component as a function of total age of the star. Assuming that at the age of 1\,Gyr the scale height is 75\,pc \citep{wegg2012,buckner2014} and a simple monotonic function fit to Fig.~\ref{fig:velocity_dispersion_vs_age_observed} (green curve), we can relate the total age of the star directly to the scale height using the $W$ velocity dispersion relation. We show our derived scale height as a function of time in Fig.~\ref{fig:scale_height}. Thus, the older the star is, the higher its scale height, and as such, older white dwarfs are less likely to be found within the 40\,pc volume for the same stellar formation history. Note that, our velocity dispersion relation for the $W$ velocity component only goes up to around 7\,Gyr. If the formation time is larger than this value, we assume that the scale height is the same as the scale height at the last available age of the observed velocity dispersion relation. Once the $x,y$- and $z$-coordinates are generated, any star that is more than 40\,pc from the Sun is rejected, and the process is repeated until a star within 40 pc is created.

An initial mass is then assigned for each star, assuming a mass distribution proportional to $M_{*}^{-2.3}$, where $M_{*}$ is the initial mass \citep{kroupa2001,kroupa2002}. By using the main sequence age relation of \cite{hurley2000}, the main sequence lifetime can be derived from its initial mass and from the assumed solar metallicity, which we set as $Z=0.0134$ \citep{asplund2009}. Note that this is the present-day photospheric metallicity of the Sun and not the "proto-Solar" interior metallicity at the moment of the formation of the star. Through the initial-to-final mass relation, IFMR, the initial mass of the star gives us the final white dwarf mass. In the default simulation we use the IMFR of \cite{elbadry2018}. We will address these assumptions further in Sect.~\ref{sec:ass_bias}.

The cooling age of the white dwarf is defined as the difference between the formation time of the initial star (total age) and the main sequence lifetime. If the resultant white dwarf cooling age is negative (including any merger delay or hastening as discussed in Sect.~\ref{sec:binary_bias}), it means that there has not been enough time for the star to evolve past the main-sequence and form a white dwarf.  Therefore, the process is restarted until a suitable white dwarf candidate is found. 

The mass, cooling age and absolute $G$ magnitude of the simulated white dwarf is based on evolutionary models from \citet{bedard2020} and appropriate model atmospheres with H- or He-composition \citep{tremblay2011,cukanovaite2021}. In our simulation we assume that 25\% of all white dwarfs have helium-dominated atmospheres, hence thin hydrogen envelopes and appropriate synthetic magnitudes for DB or DC white dwarfs, which is based on the observed number from the 40\,pc sample \citep{mccleery2020}. Other studies have found similar numbers (see, for example, \citealt{lopez2022}). At this point, the simulation has produced an absolute $G$ magnitude distribution that can be compared against real data.

\subsection{Binary system interactions}~\label{sec:binary_bias}

\begin{figure}
	\includegraphics[width=\columnwidth]{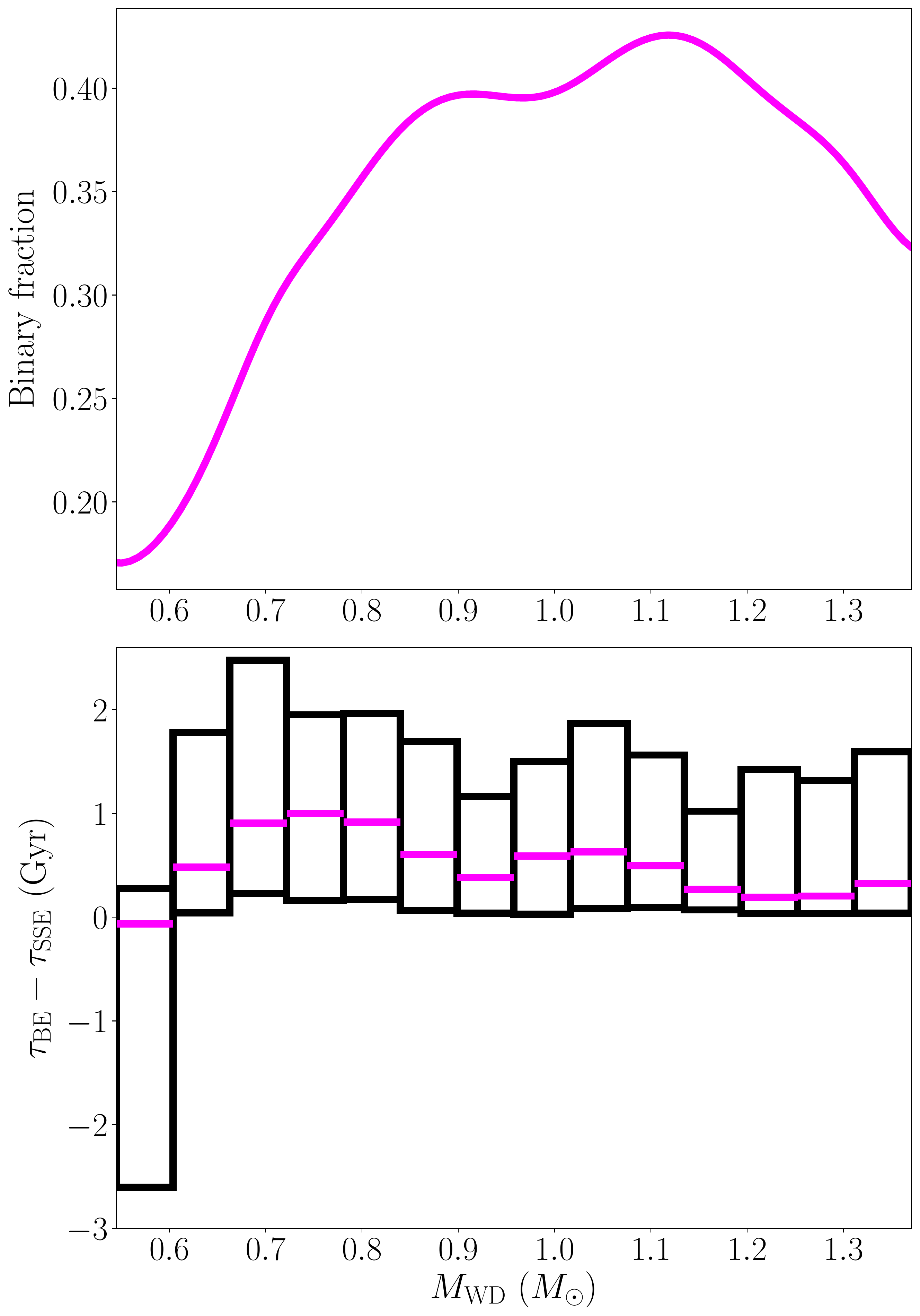}
    \caption{\textit{Top figure:} The fraction of white dwarfs that have formed via a merger as a function of the mass of the white dwarf. This is the default model from \protect\cite{temmink2020}. \textit{Bottom figure:} The time difference in Gyr between the prior history of a white dwarf assuming it has formed via binary evolution, $\tau_{\mathrm{BE}}$, and single star evolution, $\tau_{\mathrm{SSE}}$, as a function of white dwarf mass. The prior history is the time taken to form the final observed white dwarf, all the way from the formation of the original star, including any subsequent merger events in the case of binary evolution model. The pink horizontal lines indicate the median value of the time difference. The black boxes indicate the first and third quartiles of the time difference distributions. The values shown here are from the default model of \protect\cite{temmink2020}.}
    \label{fig:binary_temmink}
\end{figure}

In the simulation, we also take into account time delays which arise from binary mergers. To our knowledge, previous studies of stellar formation history involving white dwarfs did not address this delay. To take this into account in our simulation, we use the results of the binary population simulations of \cite{temmink2020}, namely the merger progenitor fractions and the time delays or speed ups due to the mergers. In particular we use the default models from \cite{temmink2020} Figs. 6 and 9, which we replot in Fig.~\ref{fig:binary_temmink}. \cite{temmink2020} employ the binary population synthesis (BPS) code SeBa \citep{portegieszwart1996,toonen2012} to simulate large numbers of single stars and binary systems. In SeBa, processes such as stellar winds, mass transfer, angular momentum loss, common envelope phases, gravitational radiation and stellar mergers are considered with the appropriate prescriptions. Details of the setup of the default model can be found in \cite{temmink2020}, but we briefly summarize the most relevant parts in the following. The initial binary fraction is assumed to be 50\%, and the authors model common envelope phases with the alpha-prescription (with~$\alpha \times \lambda = 2$), which is based on the energy budget in the binaries. The authors then consider all single stars and binary systems that eventually form a single white dwarf. The most important contributing channels can be seen in Figs. 1 and 3 of \cite{temmink2020}. Here, we use the distributions they find in populations formed after a burst in star formation. 

If any of the apparently single 40\,pc white dwarfs are formed via a binary merger, their prior history is different to the history of a white dwarf formed through single star evolution. As shown in the bottom plot of Fig.~\ref{fig:binary_temmink}, in most cases this will mean that its prior history takes a longer time, but for lower mass white dwarfs it can actually be faster. Before we can apply this time correction, we first need to know how many white dwarfs in our simulation have formed as a result of a merger. This fraction is shown in the top panel of Fig.~\ref{fig:binary_temmink}. It is apparent that more massive white dwarfs are more likely to be formed as a result of a merger, but even at lower masses such as 0.6\,\msolar, the fraction is as high as 20\% for the default model of \cite{temmink2020}. Therefore, Fig.~\ref{fig:binary_temmink} (top panel) gives the probability of a simulated white dwarf with a given mass being the product of a merger, in which case we apply a time delay based on the delay distribution shown in the bottom panel of Fig.~\ref{fig:binary_temmink}. We determine the time correction by randomly sampling from a Gaussian distribution with a mean and standard deviation determined from the quartile information (Q1, Q2 and Q3) given in Fig. 6 of \cite{temmink2020}. Note that this corresponds to the combination of a cooling delay and extended pre-white dwarf lifetime. Since the time difference is almost always positive, in most cases it will shorten the cooling age of the white dwarf for the same stellar formation history. Therefore, when compared to a model with no binary evolution, the binary-corrected model should have hotter and brighter white dwarfs for the same formation history. Binary evolution is included in our simulation by default and is used in the following analysis. 

We remind the reader that we also account for binarity by removing the majority of double degenerate candidates from our mass cut at 0.54\,\msolar\ introduced in Section\,\ref{sec:40pc}. Unresolved white dwarf and main-sequence pairs are also missing from our sample as it was constructed from a \textit{Gaia} selection of single or multiple white dwarfs \citep{gentilefusillo2021}.

\section{Determination of Stellar Formation History}~\label{sec:fitting}

\begin{figure*}
	\includegraphics[width=2\columnwidth]{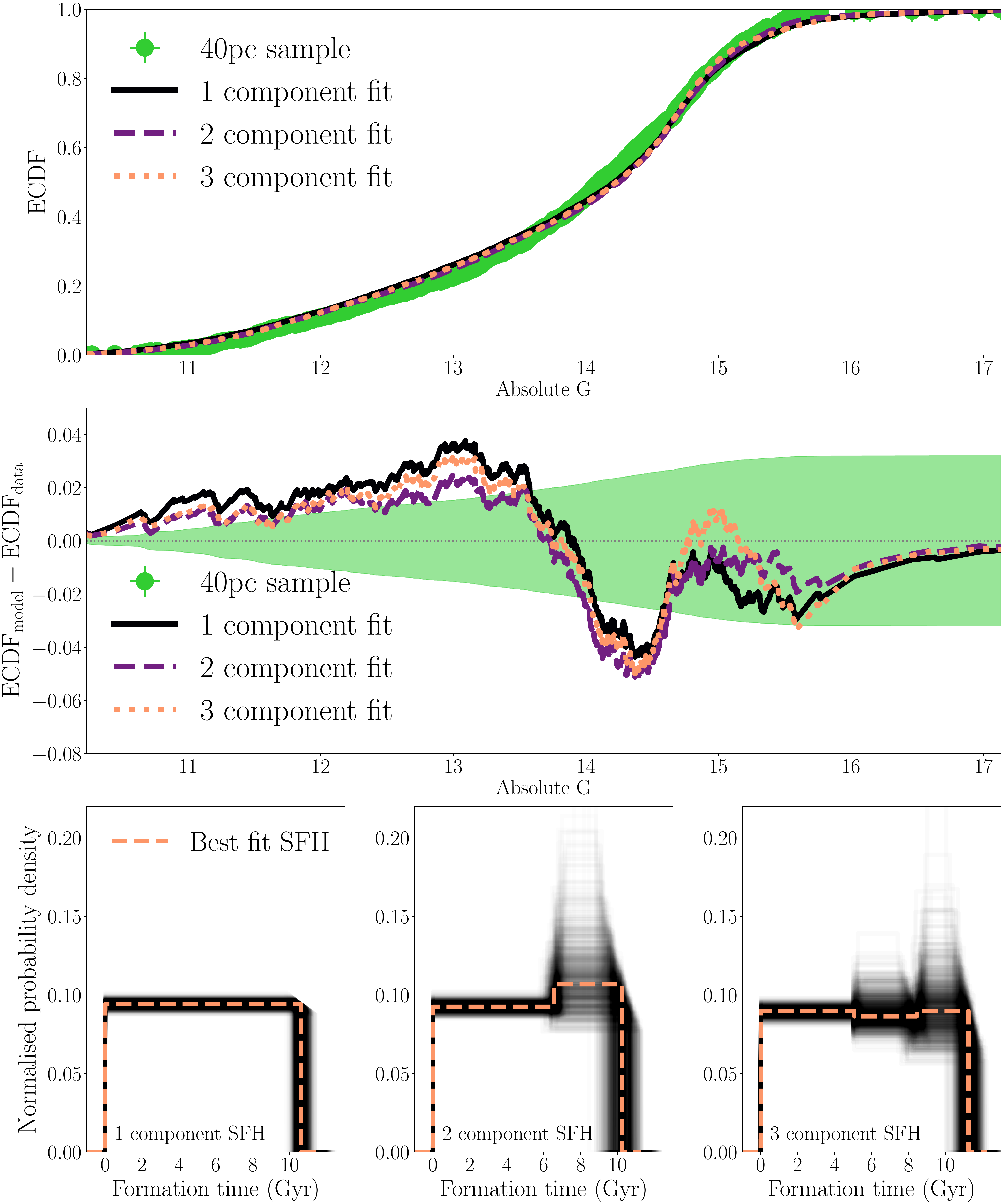}
    \caption{\textit{Top plot}: The empirical cumulative distribution function of the absolute magnitudes. The lime green error bars show the 40\,pc sample ECDF, whereas in solid black, dashed purple, and dotted orange we plot the best fit ECDFs found by assuming uniform stellar formation histories of either one, two or three Galactic components.   
    \textit{Middle plot}: The residuals from the fits compared to the observational errors. The errors are indicated as filled green area.
    \textit{Bottom left plot}: The best fitting stellar formation history assuming one Galactic component. In solid black we show all the stellar formation histories that correspond to the probability distribution function of the fitting parameter, $\tau_{\mathrm{max}}$. In dashed orange we plot the stellar formation history found from the median value of the probability distribution function of $\tau_{\mathrm{max}}$, which we define as the best fit stellar formation history found from the observed 40\,pc sample.
    \textit{Bottom centre plot}: This plot is similar to the bottom left plot, but shows the best fit two Galactic component stellar formation history.     \textit{Bottom right plot}: Same as bottom left and centre plots, but this plots shows the best fit three Galactic component stellar formation history. }
    \label{fig:ecdf}
\end{figure*}

We fit our simulated model, with variable stellar formation history, to the observed absolute $G$ magnitude distribution of the 40\,pc white dwarf sample. We choose our fitting data to be the empirical cumulative distribution function, ECDF, of absolute $G$ magnitude, which unlike the histogram shown in Fig.~\ref{fig:absG_hist} does not depend on arbitrarily chosen bin number or size. Instead, the number of bins is determined by the number of observed white dwarfs. The ECDF of the 40\,pc white dwarf sample is shown in Fig.~\ref{fig:ecdf}. The errors on the absolute $G$ magnitude are small enough to be invisible in the figure, therefore, we use propagated Poisson errors.

For a given simulation, i.e. given stellar formation history, we calculate the ECDF using all 3000 simulated white dwarfs. To directly compare our simulation to the data, we have to scale the simulated ECDF to the number of white dwarfs in our data sample, i.e., $\sim$1000. We do this by interpolating the simulated ECDF at the values of observed 40\,pc absolute $G$ magnitudes. As mentioned previously, each simulation depends on a stellar formation history, which we parameterise in two ways. In the first case, we assume that there is only one Galactic component in the simulation, and that the stellar formation history is uniform between present day (formation time of 0\,Gyr) and some maximum time, $\tau_{\mathrm{max}}$. When fitting this scenario to the observed 40\,pc sample, we only fit for the parameter $\tau_{\mathrm{max}}$. 

In the two Galactic component case, we assume that an older component starts to uniformly produce stars at a time, $\tau_{\mathrm{max, 2}}$ until the time, $\tau_{\mathrm{max, 1}}$. At this point the older component shuts down, and the younger components starts producing stars from the time $\tau_{\mathrm{max, 1}}$ until present day. In this case, we also have a parameter $N_{2}$, which describes the fraction of stars in the older component. Note that the fraction of stars in younger component will be 1-$N_{2}$. Therefore, in this case we have three fitting parameters: $N_{2}, \tau_{\mathrm{max, 1}}, \tau_{\mathrm{max, 2}}$. The three Galactic component case is similar, but the fitting parameters are $N^{'}_{2}, N^{'}_{3}, \tau^{'}_{\mathrm{max, 1}}, \tau^{'}_{\mathrm{max, 2}}, \tau^{'}_{\mathrm{max, 3}}$, where the subscript number refers to either the first, second or third Galactic component.

As an aside, note that the absolute stellar formation rate is set by the total number of white dwarfs within 40\,pc, and so is not a free parameter. Since the vast majority of local stars are M dwarfs that have not yet formed white dwarfs, we refrain from the derivation of the absolute stellar formation rate per unit volume and time. Furthermore, we do not correct for the larger velocity of halo stars, and as such we make no attempt at deriving the absolute stellar halo formation rate. This would be difficult to constrain in any case since the halo white dwarf fraction within 40\,pc is smaller than 2.5\% \citep{torres2019}.

In each case, we fit the data by minimising the reduced weighted $\chi^2$,
\begin{equation}~\label{eq:chi2}
\chi^2 = \sum_{G'=\min(G_{\mathrm{observed}})}^{G'=\max(G_{\mathrm{observed}})} \frac{\mathrm{(simulated \ ECDF}_{G'}-\mathrm{observed \ ECDF}_{G'})^2}{\mathrm{error \ on \ observed \ ECDF}_{G'}^2 \times \mathrm{dof}},
\end{equation}
where $G'$ is the index for a given absolute magnitude bin of the ECDF; and dof is degrees of freedom, defined as 
\begin{equation}
\mathrm{dof} = n_{\mathrm{data}}-(n_{\mathrm{parameters}}-1),
\end{equation} where $n_X$ is either the number of data points or number of fitted parameters. 

For any fitting procedure, the first step is to determine the initial parameters. To do this, we explore the entirety of the physical parameter space by calculating $\chi^2$ for select number of stellar formation histories. The parameters that give the smallest $\chi^2$ are then chosen as initial parameters for the minimisation. The minimisation is based on the Scipy minimize package in Python \citep{scipy2020,python3} and is combined with our custom MCMC program to find the probability distributions of the best-fitting parameters. The final best fit parameters are then taken as the median of the probability distribution with the errors quoted being the 2.5\% and 97.5\% confidence limits. When assuming a one-component stellar formation history the best fit parameter is found to be $\tau_{\mathrm{max}} = 10.6^{+0.5}_{-0.5}$ Gyr. For the two component stellar formation history the best fit parameters are $N_2=0.39^{+0.03}_{-0.04}$, $\tau_{\mathrm{max, 1}}=6.6^{+0.5}_{-0.4}$\,Gyr, and $\tau_{\mathrm{max, 2}} = 10.4^{+0.7}_{-1.1}$\,Gyr. For the three Galactic component case, the best fit parameters are $N^{'}_{2}= 0.29^{+0.02}_{-0.02}, N^{'}_{3}=0.25^{+0.01}_{-0.02}, \tau^{'}_{\mathrm{max, 1}}=5.0^{+0.3}_{-0.3}, \tau^{'}_{\mathrm{max, 2}}=8.4^{+0.5}_{-0.6}, \tau^{'}_{\mathrm{max, 3}}=11.2^{+0.7}_{-0.7}$. 

The best fits are shown visually in  Fig.~\ref{fig:ecdf}. The top panel shows the observed and model ECDFs, the middle plot shows the residuals of the fits compared to the observational errors, and the bottom plot shows the corresponding SFHs for the one, two or three Galactic component case. Note that since we use MCMC to derive the best fit parameters, we find their probability density functions. These can be converted to the probability density functions of the best fit stellar formation histories. These are shown as black lines in the bottom plot of Fig.~\ref{fig:ecdf}, where the orange dashed lines indicate the median value of the SFH, which we take to be our best fit SFH. While looking at the top and middle plots of Fig.~\ref{fig:ecdf} we can see that there are no significant differences between the fits. Even the best fit SFHs shown in the bottom are all similar, with only minor deviations from complete uniformity for the SFHs with two or three Galactic components. 
Beyond the work shown in this paper, we have tried fits with skewed Gaussians instead of uniform distributions, both with one, two and three Galactic components. These more complex stellar formation histories do provide fits on par or in slightly better agreements with the 40\,pc observations. This is not surprising, since more complex models with larger number of parameters allow for more molding of resultant ECDF. As such, better fitting to the data could be due to over-fitting. Therefore, our philosophy is to assume the simplest solution which gives a reasonable fit to the data, i.e. a uniform SFH with one Galactic component. We will show this in the following section by comparing it to other stellar formation histories from literature. 

\section{Comparison to other stellar formation histories}~\label{sec:other_sfh}

\begin{figure*}
	\includegraphics[width=2\columnwidth]{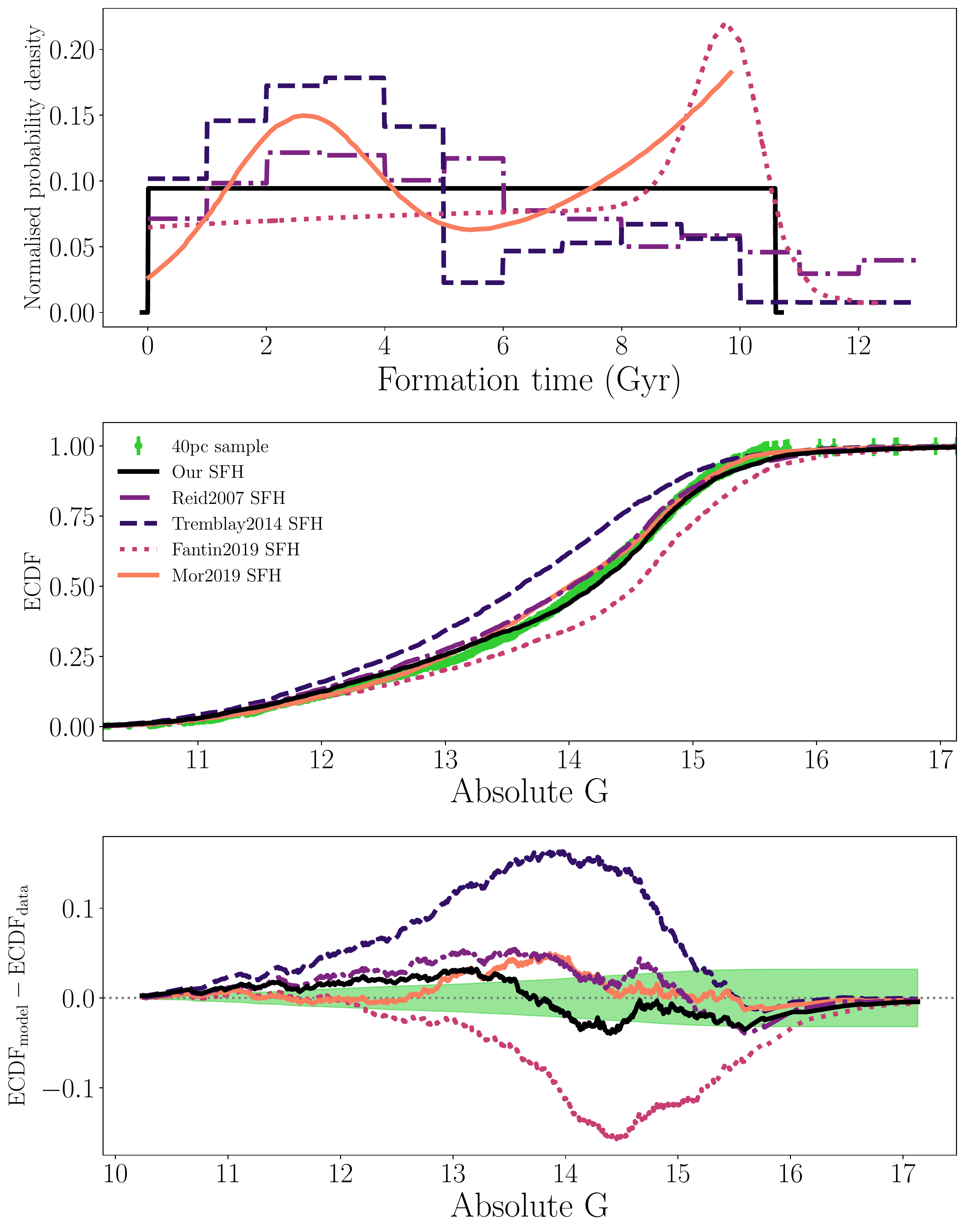}
    \caption{
    \textit{Top plot}: A comparison between our one Galactic component stellar formation history and stellar formation histories found from literature. The SFH of \protect\cite{reid2007}, \protect\cite{tremblay2014}, \protect\cite{fantin2019} and \protect\cite{mor2019} are plotted in dashed-dotted purple, dashed blue, dotted pink and solid orange, respectively. Our SFH is plotted in solid black. All stellar formation histories are normalised.
    \textit{Middle plot}: The observed ECDF plotted as green errorbars is compared to simulated ECDFs when the SFH in our simulation is changed to the SFHs of \protect\cite{reid2007}, \protect\cite{tremblay2014}, \protect\cite{fantin2019} and \protect\cite{mor2019}. These are plotted in solid black, dash-dotted purple, dashed blue, dotted pink and solid orange, respectively. 
    \textit{Bottom plot}: The residuals of the above plot.
    }
    
    \label{fig:other_stellar formation history}
\end{figure*}

In the left plot of Fig.~\ref{fig:other_stellar formation history} we compare the ECDFs found from our simulations with different stellar formation histories. These histories were taken from \cite{reid2007}, \cite{tremblay2014}, \cite{fantin2019} and \cite{mor2019} studies, and are shown in the right plot of Fig.~\ref{fig:other_stellar formation history}. The stellar formation histories of \cite{reid2007} and \cite{mor2019} are based on main sequence stars, whereas \cite{tremblay2014} and \cite{fantin2019} are based on white dwarfs. The \cite{mor2019} and \cite{fantin2019} results are based on \gaia~data. We chose these particular SFHs as they have different shapes, specifically they show peaks at different times. 

From this comparison we find that our uniform SFH and the SFHs from \cite{reid2007} and \cite{mor2019} fit the data, whereas the SFHs from \cite{tremblay2014} and \cite{fantin2019} do not agree with the data and can be excluded. The data from the 40\,pc sample favours a more flat SFH such as ours or \cite{reid2007}, or a SFH with two peaks such as the SFH of \cite{mor2019}. 

\begin{figure*}
	\includegraphics[width=2\columnwidth]{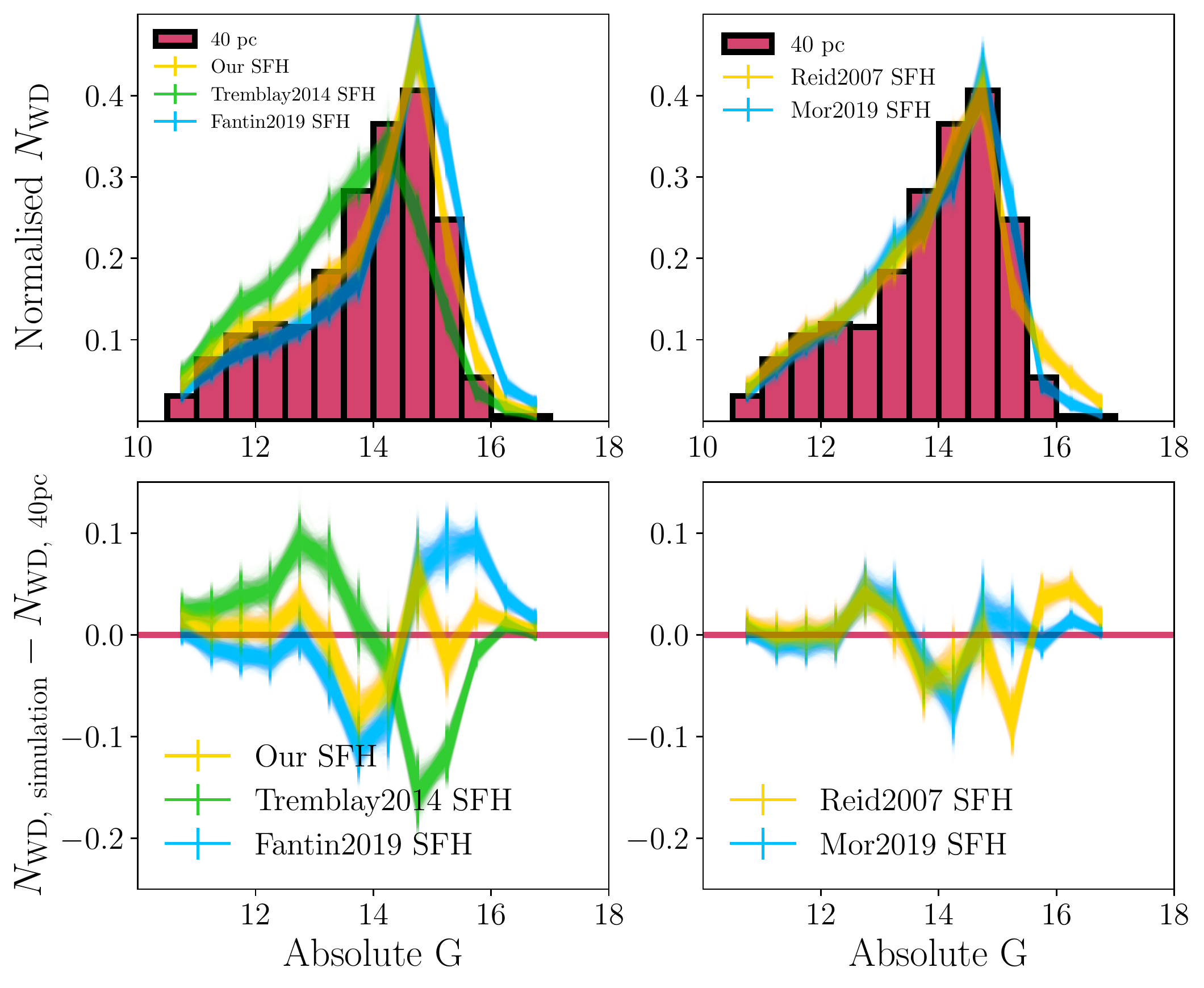}
    \caption{\textit{Top left plot}: A histogram of the absolute $G$ magnitudes for the 40\,pc sample is plotted in solid pink. The histogram is normalised to unit area. The absolute $G$ magnitude histogram found from the best fit of our one-component simulation is plotted in yellow. In green and blue we plot the absolute $G$ magnitudes from our simulation relying on the stellar formation history of \protect\cite{tremblay2014} and \protect\cite{fantin2019}, respectively. The residuals of this comparison are shown in the bottom left plot, with the solid pink line indicating perfect agreement.
    \textit{Top right plot}: Same as the top left plot, but in yellow and blue we plot the histograms found from our simulation with the stellar formation histories taken from \protect\cite{reid2007} and \protect\cite{mor2019}. The bottom right plot shows the corresponding residuals. 
    }
    
    \label{fig:other_sfh_abs_g_hist}
\end{figure*}

In Fig.~\ref{fig:other_sfh_abs_g_hist} we show a comparison of the absolute $G$ magnitude histograms from the observed 40\,pc white dwarf sample and simulations with published stellar formation histories of \cite{reid2007}, \cite{tremblay2014}, \cite{fantin2019} and \cite{mor2019}. The residuals of the comparisons are shown in the bottom plots of Fig.~\ref{fig:other_sfh_abs_g_hist}. As for the case of the empirical cumulative distribution function of absolute $G$ magnitudes, the best agreement is with our own stellar formation history, as well as those of \cite{reid2007} and \cite{mor2019}.

In the following, we further test the robustness of our stellar formation history by assessing how various assumptions made in our models affect the best fit.
\section{Theoretical assumptions}~\label{sec:ass_bias}

Assumptions have been made when building our simulation. These include complex assumptions based on difficult physical considerations, such as the relation between the initial and final mass of a star as it becomes a white dwarf. In the following, our aim is to assess whether the assumptions made in our simulations have a significant effect on the derived stellar formation history. We do this by changing one assumption at a time and assessing how this change affects the best-fit $\chi^2$ between data and simulations. Note that the value of $\chi^2$ can indicate whether a given hypothesis is in agreement with the data. For this analysis we proceed with a one-component, uniform stellar formation history as we have shown that it gives a reasonable fit to the data. 
 
\subsection{Metallicity}~\label{sec:metal}

In the simulation, the assumed metallicity has a direct effect on the derived main sequence lifetime, which depends both on the metallicity and on the main sequence mass. In general, for main sequence masses below $\approx6$\,\msolar, metallicities smaller than the solar metallicity result in shorter main sequence lifetimes. If the formation time does not change, i.e. we use the same stellar formation history, then the resulting white dwarfs will have longer cooling ages and dimmer magnitudes. At around 6\,\msolar~and above, the metallicity has the opposite effect on the main sequence lifetime. Due to the initial mass function, most main sequence stars in our simulation will have lower masses, i.e., below 6\,\msolar. Thus, the reversal of the effect of metallicity above 6\,\msolar~is not significant for our simulation.

When trying to fit data with our simulation, the effect of metallicity on our SFH and therefore the derived value of the parameter $\tau_{\mathrm{max}}$ is the following: A simulation with larger assumed metallicity will produce more bright white dwarfs than a simulation with low metallicity if the stellar formation history stays the same. Therefore, to compensate for this, the fitted value of $\tau_{\mathrm{max}}$ should be larger for the simulation with larger metallicity in order to produce enough dim white dwarfs. 
Similarly, it could change the shape of the fitted SFH from uniform to a peak at later times, to over-compensate for too many simulated bright white dwarfs. 
We run three additional types of simulations with different assumed metallicities: $0.2+Z_{\odot}$, $Z_{\odot}/10$ and $Z_{\odot}/1000$, where $Z_{\odot}$ is the solar metallicity equal to 0.0134, used in the original simulation. These particular choices of metallicity are extreme and have been chosen to test the robustness of our derived stellar formation history. The metallicity of $0.2+Z_{\odot}$ represents metal-rich stars, which have metallicity around 0.2 dex higher than the solar metallicity \citep{feltzing2001}. The metallicities of $Z_{\odot}/10$ and $Z_{\odot}/1000$ represent the range of the metallicities found in metal-poor population II stars. Note that for this paper we assume that the metallicity is constant within the 40\,pc sample, as \cite{rebassa2021} showed from white dwarfs with wide stellar companions and within 500\,pc of the Sun that there is no evidence of metallicity changing as a function of formation time.

The 40\,pc sample ECDF is fitted with the three types of simulations using the MCMC method with 1000 walkers. Each walker finds a best fit value of $\tau_{\mathrm{max}}$ and the corresponding $\chi^2$ value. Thus, for each simulation we find a probability density function of both $\chi^2$ and $\tau_{\mathrm{max}}$. In Table~\ref{tab:ass} we list the median values of $\chi^2$ and $\tau_{\mathrm{max}}$, and as their errors we show the confidence limits of 2.5\% and 97.5\% of their distributions. As expected, the increased metallicity results in a larger value of $\tau_{\mathrm{max}}$ when compared to other metallicities and the results from the original simulation. The $\chi^2$ values stay close to 1 in all cases and thus indicate that different values of metallicity still allow the uniform SFH to agree with the data, without needing a more peaked SFH.

\renewcommand{\arraystretch}{1.5}

\begin{table*}
	\centering
	\caption{The best fit $\chi^2$ and $\tau_{\mathrm{max}}$ found from simulations where a specific assumption is changed. The first column indicates which assumption is changed, with Metallicity, IMF, $\tau_{\mathrm{MS}}$, IFMR, Velocity correction, Binary correction, He-fraction, WD cooling model referring to tests with changes in either the metallicity, initial mass function, main sequence lifetime, initial-to-final mass relation, the velocity correction, binary correction, the fraction of helium-dominated atmosphere white dwarfs, the white cooling model, or the 40\,pc sample from mass correction and double degenerate removal, respectively. The value of the assumption is listed in the second column, with more information provided in the text. Third column lists the best fit $\chi^2$ found from 1000 walkers, with the errors indicating the confidence limits of 2.5\% and 97.5\% of $\chi^2$ probability density functions. The last column is the same but for the parameter $\tau_{\mathrm{max}}$, which uniquely defines the uniform, one Galactic component SFH.}
	\label{tab:ass}
	\begin{tabular}{|l|l|c|c|} 
 \hline
 Simulation & Assumption & $\chi^2$ & $\tau_{\mathrm{max}}$ (Gyr) \\ 
 \hline
 Original &  & $1.4^{+0.8}_{-0.7}$ & $10.6^{+0.5}_{-0.5}$ \\
 \hline
 Metallicity & $Z_{\odot}+0.2$ & $1.1^{+0.8}_{-0.5}$ & $11.0^{+0.5}_{-0.7}$ \\
  & $Z_{\odot}/10$ & $1.3^{+0.8}_{-0.6}$ & $10.0^{+0.4}_{-0.4}$ \\ 
  & $Z_{\odot}/1000$ & $1.5^{+0.9}_{-0.6}$ & $9.6^{+0.4}_{-0.4}$ \\ 
 \hline
 IMF & Eq.~\ref{eq:imf_perr} (steeper) & $2.6^{+2.0}_{-1.0}$ & $11.7^{+0.7}_{-1.1}$\\
  & Eq.~\ref{eq:imf_merr} (shallower) & $0.6^{+0.5}_{-0.3}$ & $9.6^{+0.4}_{-0.4}$ \\
  \hline
  $\tau_{\mathrm{MS}}$ & $(1-0.048)*\tau_{\mathrm{MS}}$ & $1.4^{+0.9}_{-0.6}$ & $10.5^{+0.6}_{-0.2}$ \\
   & $(1+0.048)*\tau_{\mathrm{MS}}$ & $1.4^{+0.9}_{-0.6}$ & $10.6^{+0.6}_{-0.5}$ \\
   \hline
   IFMR & \cite{cummings2018} & $1.0^{+0.9}_{-0.5}$ & $11.1^{+0.6}_{-0.6}$ \\
   & \cite{andrews2015} & $1.5^{+0.9}_{-0.7}$ & $9.6^{+0.5}_{-0.4}$ \\
   \hline
   Velocity correction & Not included & $1.5^{+0.9}_{-0.7}$ & $10.6^{+0.6}_{-0.5}$ \\
   \hline
   Binary correction & Not included & $1.5^{+0.9}_{-0.7}$& $10.4^{+0.5}_{-0.6}$ \\
   \hline
   He-fraction & 0\% & $1.7^{+0.9}_{-0.7}$  & $10.6^{+0.5}_{-0.5}$ \\
  & 50\% & $1.4^{+0.9}_{-0.7}$  & $10.6^{+0.5}_{-0.5}$ \\
  \hline
  WD cooling model & LPCODE & $4.6^{+1.6}_{-1.2}$&  $9.7^{+0.6}_{-0.6}$\\
  \hline
  40\,pc mass correction & No correction &  $1.6^{+0.8}_{-0.6}$ & $8.6^{+0.5}_{-0.5}$\\
and double degenerate removal & No correction and no removal & $1.1^{+0.8}_{-0.5}$ & $10.3^{+0.5}_{-0.5}$ \\
\hline
 \end{tabular}
\end{table*}

\begin{figure*}
	\includegraphics[width=1.8\columnwidth]{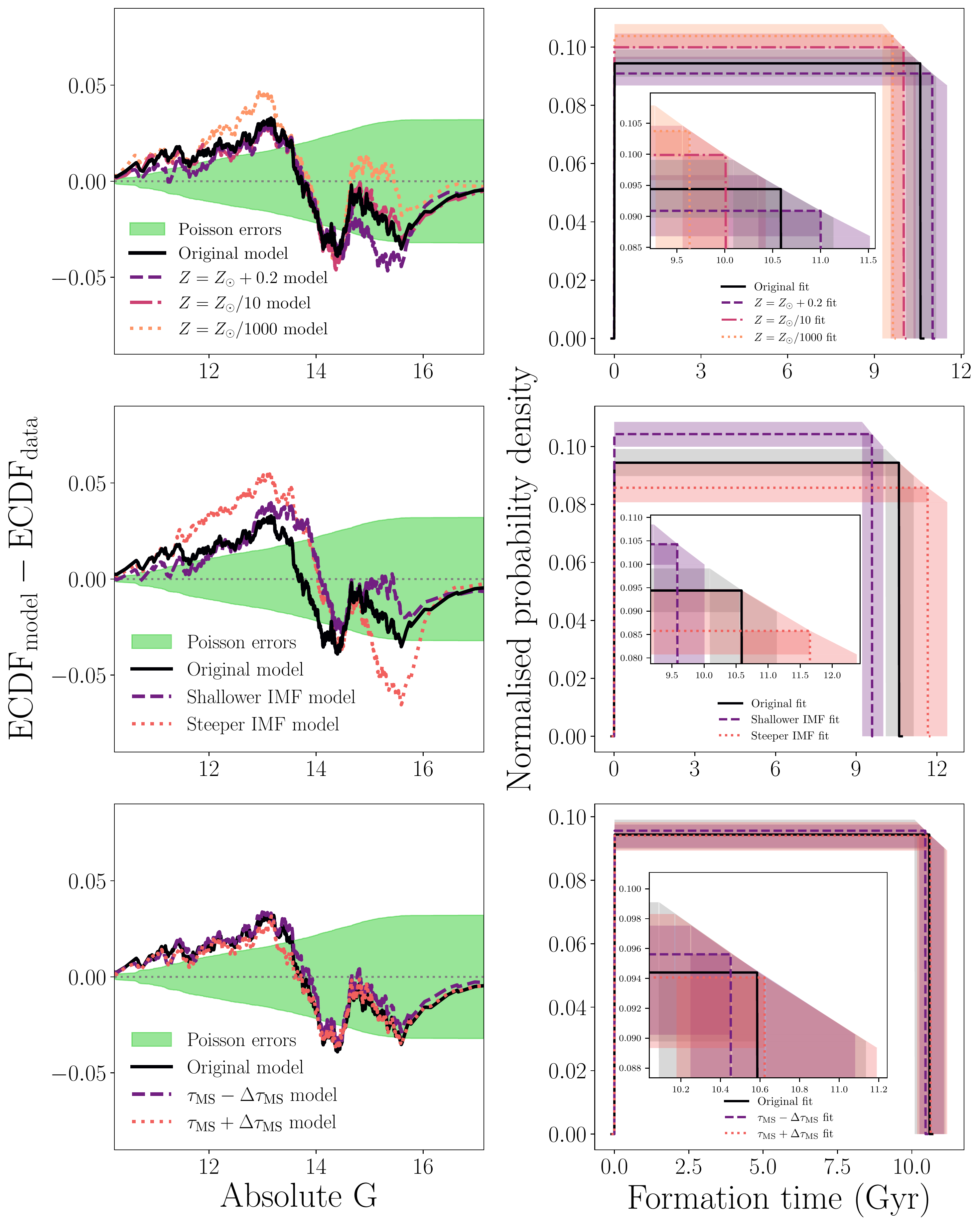}
    \caption{\textit{Left plots}: The residuals from the fits of different simulations to the 40\,pc data. In all left plots the solid black line indicates the residuals between the original simulation and the data. The green filled area denotes the propagated Poisson errors. Top, centre and bottom left plots show the residuals when the metallicity, the IMF, and the main sequence lifetimes are changed, respectively. The legends indicate which coloured lines represent a given assumption and its chosen value. \textit{Right plots}: The one-component, uniform stellar formation histories found from fitting the 40\,pc sample with simulations where a single assumption is changed. Each right plot corresponds to the left plot on the same row. The right plots shows the full stellar formation history, where the formation time of zero Gyr means present day.  The legends indicate the best fit stellar formation histories and which assumption they correspond to. The corresponding coloured areas represent the 2.5\% and 97.5\% confidence limits of the derived stellar formation history when the relevant assumption is changed. For clearer illustration of the differences in the derived stellar formation history, the inset plot shows the zoomed in area near the best fit formation time of $\tau_{\mathrm{max}}$.}
    \label{fig:met_imf_taums}
\end{figure*}

In the top left plot of Fig.~\ref{fig:met_imf_taums} we show the residuals of the fits and how they compare to the observed propagated Poisson errors, which are plotted as filled green area. It is clear that all simulations give reasonable fits, but the median value of $\chi^2$ is smallest for the simulation with $Z_{\odot}+0.02$. However, looking at the top left plot of Fig.~\ref{fig:met_imf_taums} it is clear that the higher metallicity does not describe dimmer magnitudes as well as other simulations, especially the $Z_{\odot}/1000$ simulation. 

In the top right plot of Fig.~\ref{fig:met_imf_taums} we show the best fit stellar formation history for each type of simulation, alongside the confidence limits of 2.5\% and 97.5\%, which are shown as correspondingly coloured areas. There is a significant overlap between the original simulation and the other three types of simulations, suggesting that the metallicity has a small effect on the derived stellar formation history within the framework of one Galactic component uniform SFH. If we take into account that the chosen metallicities are extreme choices, this conclusion becomes more robust, as it is unlikely that within 40\,pc the metallicity could change so drastically from solar metallicity. We conclude that the assumed metallicity does not change our conclusion that a one-component Galactic uniform SFH fits the 40\,pc data well.

\subsection{Initial mass function}~\label{sec:imf}

In the original simulation we use the initial mass function (IMF) from \cite{kroupa2001}, namely, $\rho(M) \propto M^{-2.3}$, where $M$ is the initial mass of the star and $\rho(M)$ is the initial mass density function. \cite{kroupa2001} provides errors for the power index of the function, which depends on the mass of the star. In Eqs.~\ref{eq:imf_perr} and~\ref{eq:imf_merr} we show the resultant initial mass density functions when the errors are either added to or subtracted from the power index of the \cite{kroupa2001} IMF. 
\begin{equation}~\label{eq:imf_perr}
\begin{gathered}
\rho(M) \propto M^{-2.6} \mathrm{ \ for \ } M < 1M_{\odot}, \\
\rho(M) \propto M^{-3.0} \mathrm{ \ for \ } M \geq 1M_{\odot},
\end{gathered}
\end{equation}

\begin{equation}~\label{eq:imf_merr}
\begin{gathered}
\rho(M) \propto M^{-2.0} \mathrm{ \ for \ } M < 1M_{\odot}, \\
\rho(M) \propto M^{-1.6} \mathrm{ \ for \ } M \geq 1M_{\odot}.
\end{gathered}
\end{equation}
Eq.~\ref{eq:imf_perr} is steeper than the original IMF used, whereas Eq.~\ref{eq:imf_merr} is shallower. To test whether the IMF affects the accuracy of the derived stellar formation history, we run two additional simulations, where we replace the \cite{kroupa2001} IMF with the Eqs.~\ref{eq:imf_perr} and~\ref{eq:imf_merr}. Steeper IMF means that more low mass main sequence stars will be created. Therefore, main sequence lifetimes will be on average longer than in the original simulation, and if we assume the same stellar formation history and thus formation times, the resultant white dwarf cooling ages will be shorter. This means that for the same stellar formation history, the simulated ECDF will have more white dwarfs at brighter magnitudes. This is a similar effect to increasing metallicity as described in Sect.~\ref{sec:metal}. The opposite is true for the shallower IMF from Eq.~\ref{eq:imf_merr}.

In Table~\ref{tab:ass} we show the best fit values of $\chi^2$ and $\tau_{\mathrm{max}}$. The much larger value of $\chi^2$ for the simulation with steeper IMF indicates that for this assumption the uniform SFH does not agree with the data within one standard error. For the shallower IMF, we get a better fit than compared to the original model. These two conclusions are supported by the central row left plot of Fig.~\ref{fig:met_imf_taums}. The central right plot of Fig.~\ref{fig:met_imf_taums} agrees with our expectations of steeper IMF producing more bright white dwarfs and therefore resulting in larger $\tau_{\mathrm{max}}$ to produce enough dim white dwarfs to match observations when compared to the original simulation. Both figures show that the chosen IMF has a moderate effect on the derived stellar formation history.

\subsection{Main sequence age function}

In the original simulation, the main sequence lifetime is found from the age function of \cite{hurley2000} which has an error of 4.8\%. To test whether the main sequence age function has a significant effect on the derived stellar formation history, we run two additional simulations where the error of 4.8\% is either added to or subtracted from the age given by the \cite{hurley2000} relation. When compared to the original simulation, an increase in the main sequence lifetime will result in shorter white dwarf cooling ages, since the formation time remains the same. Therefore, such a simulation will have more white dwarfs at brighter magnitudes in the resultant ECDF. Again, this is similar to increasing the metallicity (Sect.~\ref{sec:metal}) or having a steeper IMF (Sect.~\ref{sec:imf}). The opposite is expected for the simulation where the error is taken away from the main sequence lifetime. 

Table~\ref{tab:ass} shows the best fit $\chi^2$ and $\tau_{\mathrm{max}}$ when changing the main sequence lifetime. Additionally, the bottom row of Fig.~\ref{fig:met_imf_taums} visually illustrates the results of this test. The best fit values of $\chi^2$ and $\tau_{\mathrm{max}}$ with either test agree with the original simulation. The bottom left plot shows that visually the simulated ECDFs from the three types of simulations look identical. The probability distributions shown in centre left plot are also almost identical. It is clear that the effect on the resultant ECDF is insignificant, therefore, we proceed with using the original main sequence lifetime relation from \cite{hurley2000}, and do not investigate this assumption any further.

\subsection{Initial to final mass relation}

To test whether the IFMR has a significant effect on the derived stellar formation history we run two additional simulations which replace the \cite{elbadry2018} IFMR with the \cite{andrews2015} or \cite{cummings2018} relations, respectively. We choose these two particular IFMRs because they are all relatively recent, but also give significantly different final masses of white dwarfs. In most cases, the \cite{cummings2018} relation results in higher white dwarf masses than the \cite{elbadry2018} relation, whereas the \cite{andrews2015} relations always gives smaller white dwarf masses than the \cite{elbadry2018} relation. Note that the simulated white dwarf cooling age is not directly changed by the IFMR, since the main sequence lifetime and the stellar formation history remain the same. As per white dwarf cooling tables and model atmospheres, the white dwarf mass has a direct effect on the \gaia~$G$ magnitude. In general, an increase in white dwarf mass results in dimmer magnitudes for all white dwarf cooling ages below $13$\,Gyr. Therefore, the simulation with \cite{cummings2018} IFMR should produce more dim white dwarfs than the simulation with \cite{elbadry2018} IFMR. This is similar to what we observed with previous assumptions.

\begin{figure*}
	\includegraphics[width=2\columnwidth]{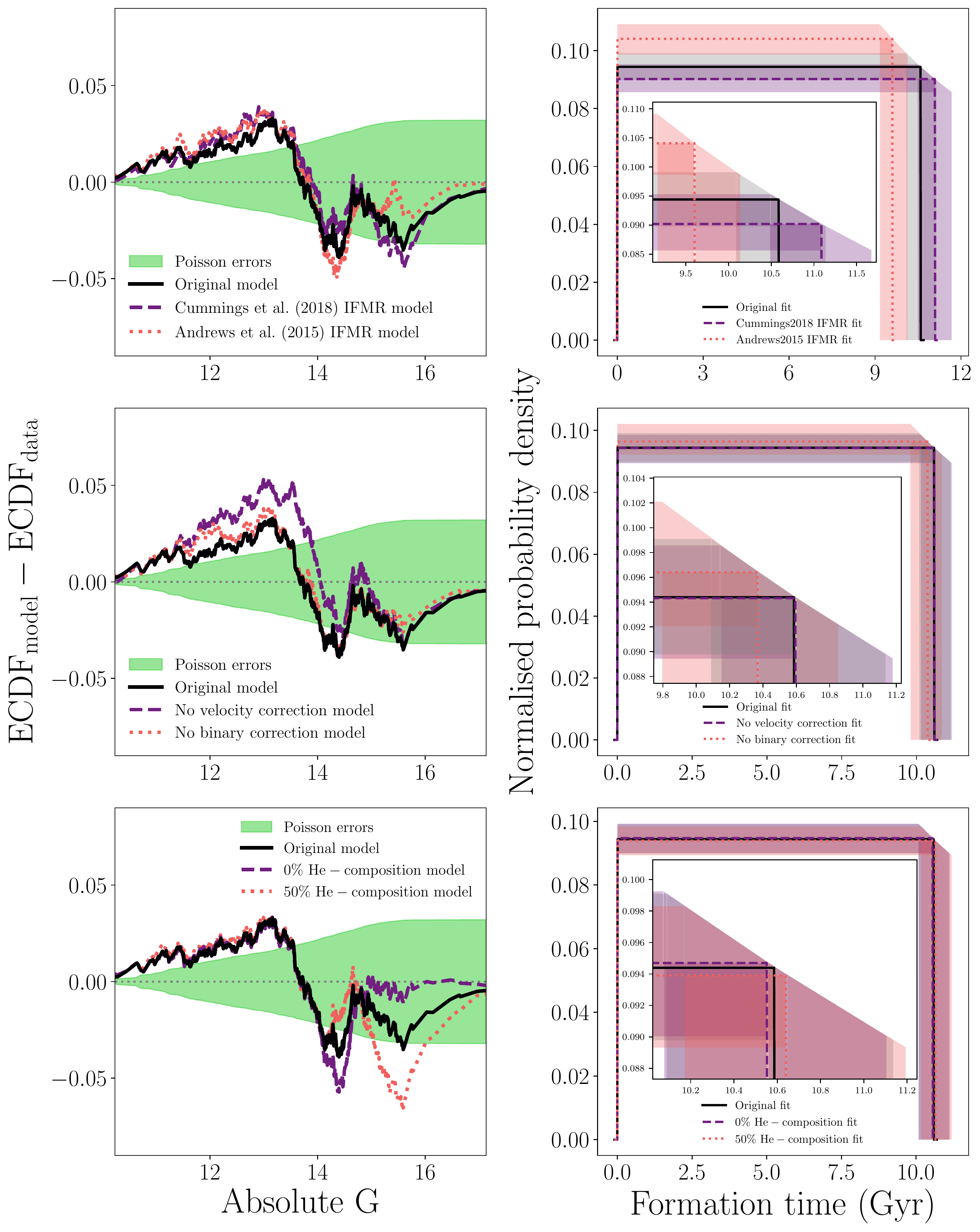}
    \caption{This figure is similar to Fig.~\ref{fig:met_imf_taums}, but here we plot the results of the tests where the initial to final mass relation is changed (top row); where the velocity correction is removed (centre row); where the binary correction is removed (centre row); and where the fraction of He-atmosphere white dwarfs is changed (bottom row).}
    \label{fig:ifmr_vel_bin_db}
\end{figure*}

Table~\ref{tab:ass} and the top row of Fig.~\ref{fig:ifmr_vel_bin_db} show the results of these tests. The $\chi^2$ between the different types of simulations does not change significantly, therefore all three simulations are broadly consistent with the observed ECDF. The simulation with \cite{cummings2018} IFMR results in larger $\tau_{\mathrm{max}}$ than the original and the \cite{andrews2015} simulations. We note that the effect of the IFMR is similar to what we observed with changing the initial mass function, except it is less significant. Both these assumptions affect the simulated masses of white dwarfs, hence these two parameters may be degenerate in determining the stellar formation history. 

\subsection{Velocity correction}

As discussed in Sect.~\ref{sec:vel_bias} and Sect.~\ref{sec:numerical}, by default our simulation includes the age-velocity dispersion relation to compensate for a bias against high-velocity objects in the 40\,pc data set. To see if this correction has a significant effect on the derived stellar formation history, we create a simulation which does not include the age-velocity dispersion relation. Instead, it assumes a constant scale height equal to 250\,pc. This values does not particularly matter because it only changes the absolute Galactic stellar formation rate which is not directly constrained in this work. This new simulation is then fitted to the observed 40\,pc data. The best fit $\chi^2$ and $\tau_{\mathrm{max}}$ are shown in Table~\ref{tab:ass}. These are in agreement with the values found from the original simulation where the age-velocity dispersion is included. The centre row of Fig.~\ref{fig:ifmr_vel_bin_db} shows the residuals and the probability distribution of stellar formation history for this test. As can be seen, there is no significant difference between the derived stellar formation histories, indicating that the velocity dispersion correction does not affect the derived stellar formation history significantly. This is in part because the correlation between white dwarf total age and absolute $G$ magnitude is weak, as the other important parameter, the progenitor mass, is left unchanged in this test simulation.

\subsection{Binary correction}

Sect.~\ref{sec:binary_bias} has described how the time delay correction arising from binary mergers is included in our simulation. In this section we assess whether this effect is significant for the derived stellar formation history. To do this, a simulation is created where the binary time delay correction from \citealt{temmink2020} is not included. As described in Sect.~\ref{sec:binary_bias} in most cases the inclusion of a binary correction means that more bright white dwarfs are produced for the same stellar formation history. Thus, when fitting the 40\,pc sample, a simulation with no binary correction would have to produce less dim white dwarfs which is done by decreasing $\tau_{\mathrm{max}}$ of the fitted stellar formation history. The results are shown in Table~\ref{tab:ass} and Fig.~\ref{fig:ifmr_vel_bin_db}. From the $\chi^2$ information it is clear that the exclusion of merger time delays does not change the best fit ECDF by more than the Poisson errors. 

\subsection{Fraction of He-atmosphere white dwarfs}

In the original simulation, the fraction of He-atmosphere white dwarfs has been set as 25\% based on the 40\,pc sample. To test whether this particular choice has a significant effect on the derived stellar formation history, we run two additional simulations where the fraction is set to 0\% and to 50\%. In general, the effect of having more He-atmosphere white dwarfs is highly non-linear, as for some white dwarf cooling ages and masses the $G$ magnitudes are dimmer than H-atmosphere white dwarfs, whereas for others the $G$ magnitudes are brighter.

In Table~\ref{tab:ass} and bottom row of Fig.~\ref{fig:ifmr_vel_bin_db} we show the results of this test. It is clear that the particular choice of He-atmosphere white dwarf fraction has a small effect on the ECDF or the derived SFH. However, we must note that the observed ECDF is better fitted at dimmer magnitudes with a 0\% He-fraction. This is an interesting result as it suggests that having a well characterised spectroscopic sample with precise atmospheric parameters and chemical compositions may be as important as finding the right IMF and IFMR for fitting the white dwarf absolute magnitude distribution. 

\subsection{White dwarf cooling tables}

To test whether the white dwarf cooling tables have an effect on the derived stellar formation history, we run a simulation with cooling tables generated from the LPCODE stellar evolutionary code \citep{althaus2005,althaus2010b,althaus2012,rohrmann2012,althaus2013,salaris2013,camisassa2016,miller2016,camisassa2017,rohrmann2018,camisassa2019}.
In general, the cooling tables directly affect the \gaia~$G$ magnitude and the differences between the Montr\'eal tables and the LPCODE tables are complex. However, if one table was to give consistently dimmer $G$ magnitudes, then the best fit values of $\tau_{\mathrm{max}}$ would decrease when fitting the same sample of data. For the same white dwarf mass and cooling ages, we find that the LPCODE results in dimmer magnitudes. 

\begin{figure*}
	\includegraphics[width=2\columnwidth]{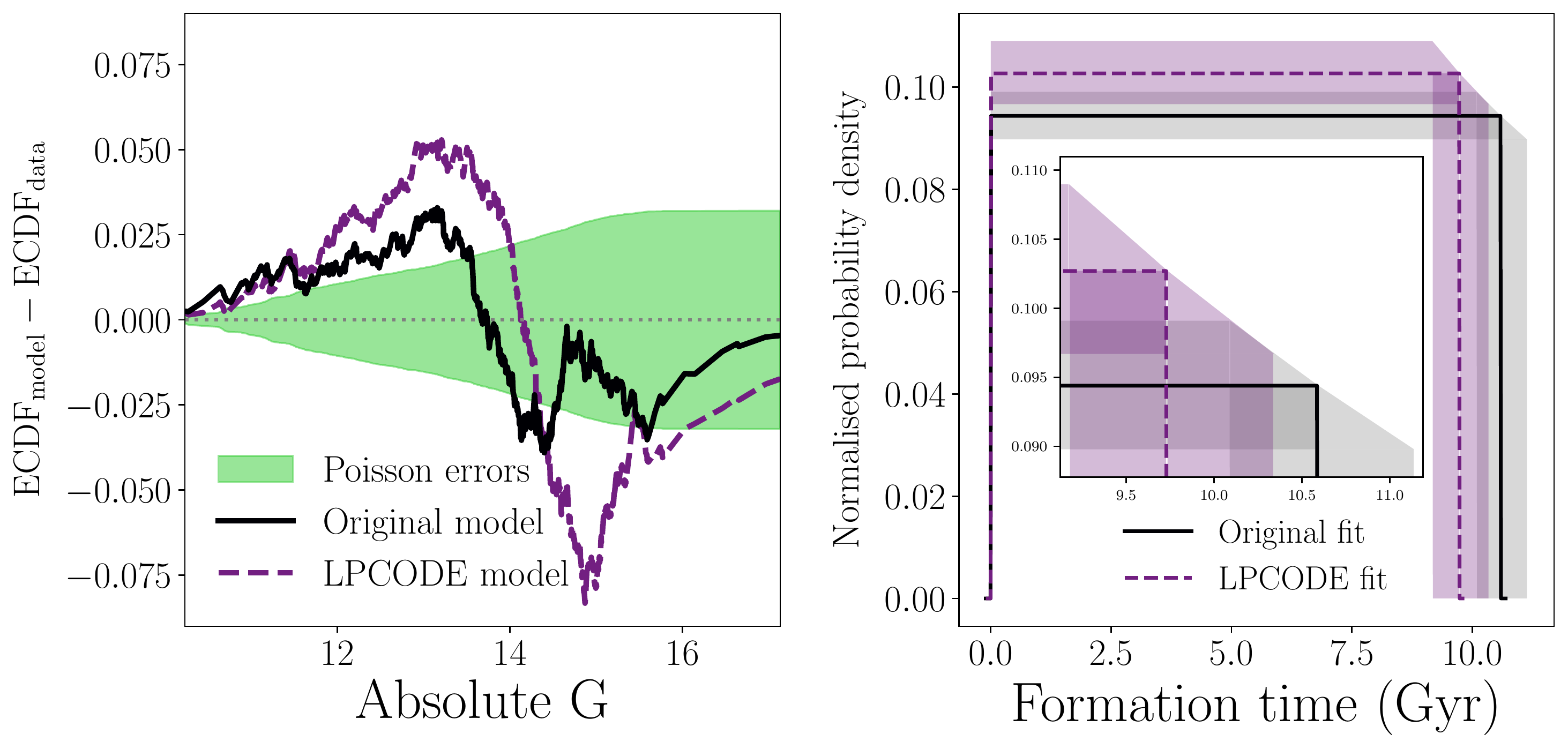}
    \caption{This figure is similar to Figs.~\ref{fig:met_imf_taums} and~\ref{fig:ifmr_vel_bin_db}, but here we plot the results of the tests where the white dwarf cooling tables are changed to those generated by the LPCODE stellar evolutionary code.}
    \label{fig:argentina}
\end{figure*}

Table~\ref{tab:ass} and Fig.~\ref{fig:argentina} show the best fit values of $\chi^2$ and $\tau_{\mathrm{max}}$ for this test. Based on $\chi^2$ this assumption disagrees the most with the hypothesis of a uniform, one Galactic component SFH. Therefore, a more complex SFH would be needed for the observed 40\,pc sample if the LPCODE is more accurate. However, it cannot be much different from a uniform distribution, such as the SFH of \cite{tremblay2014} or \cite{fantin2019}, as the residuals for those are much larger (see Fig.~\ref{fig:other_stellar formation history}). 

\subsection{Mass correction and double degenerate white dwarf removal}

{In this section, we switch from changing the different assumptions in the simulation, and instead look at the observed 40\,pc sample. In the previous sections, we modified the original 40\,pc white dwarf sample from \cite{mccleery2020} and \cite{obrien2022} by applying a mass correction defined in Table~\ref{tab:mass_corr} for the low-mass problem of cool white dwarfs (see Sect.~\ref{sec:40pc}), and then by removing any white dwarf whose mass was below the mass limit from single star evolution, i.e. double degenerate candidates. A question arises whether this modification of the 40\,pc sample has an effect on the derived stellar formation history. Therefore, to test this, we look at the following three scenarios:
\begin{enumerate}
  \item The 40\,pc sample used in all previous sections and described above, i.e. the masses of white dwarfs have been corrected and double degenerates have been removed. For this test we will refer to this sample as the "Corrected 40\,pc sample".
  \item The "Uncorrected 40\,pc sample" refers to the 40\,pc sample where mass correction was not applied, but any white dwarf whose mass is below the limit from single star evolution was removed.
  \item  The last sample is the "Whole 40\,pc sample" of \cite{mccleery2020} and \cite{obrien2022}, where no mass correction is applied and no white dwarf has been removed.
\end{enumerate}

In Fig.~\ref{fig:mass_corr_tests} we show the empirical cumulative distributions and the histograms of the absolute $G$ magnitude for the three 40\,pc samples. Over plotted on the histogram are the Poissonian errors for the "Corrected 40\,pc sample" as a reference to judge whether the other two samples are significantly different. Note that these errors are also plotted in Fig.~\ref{fig:absG_hist}.

It is clear from the top plot that the ECDFs of the "Uncorrected 40\,pc sample" is significantly different from the other two samples. As the positive mass correction is not applied, the removal of white dwarfs with a mass below the mass limit results in a larger number of white dwarfs removed when compared to the "Corrected sample", i.e. 280 white dwarfs are removed as opposed to 110 which were removed to create the "Corrected 40\,pc sample". We note that this extra removal is likely spurious, i.e. not removing true double degenerates. Nevertheless, these lower mass white dwarfs are cooler and older, and thus their removal in the "Uncorrected sample" results in a histogram with more bright white dwarfs. Predictably, this should mean a SFH which creates white dwarfs closer to present time.

\begin{figure*}
	\includegraphics[width=2\columnwidth]{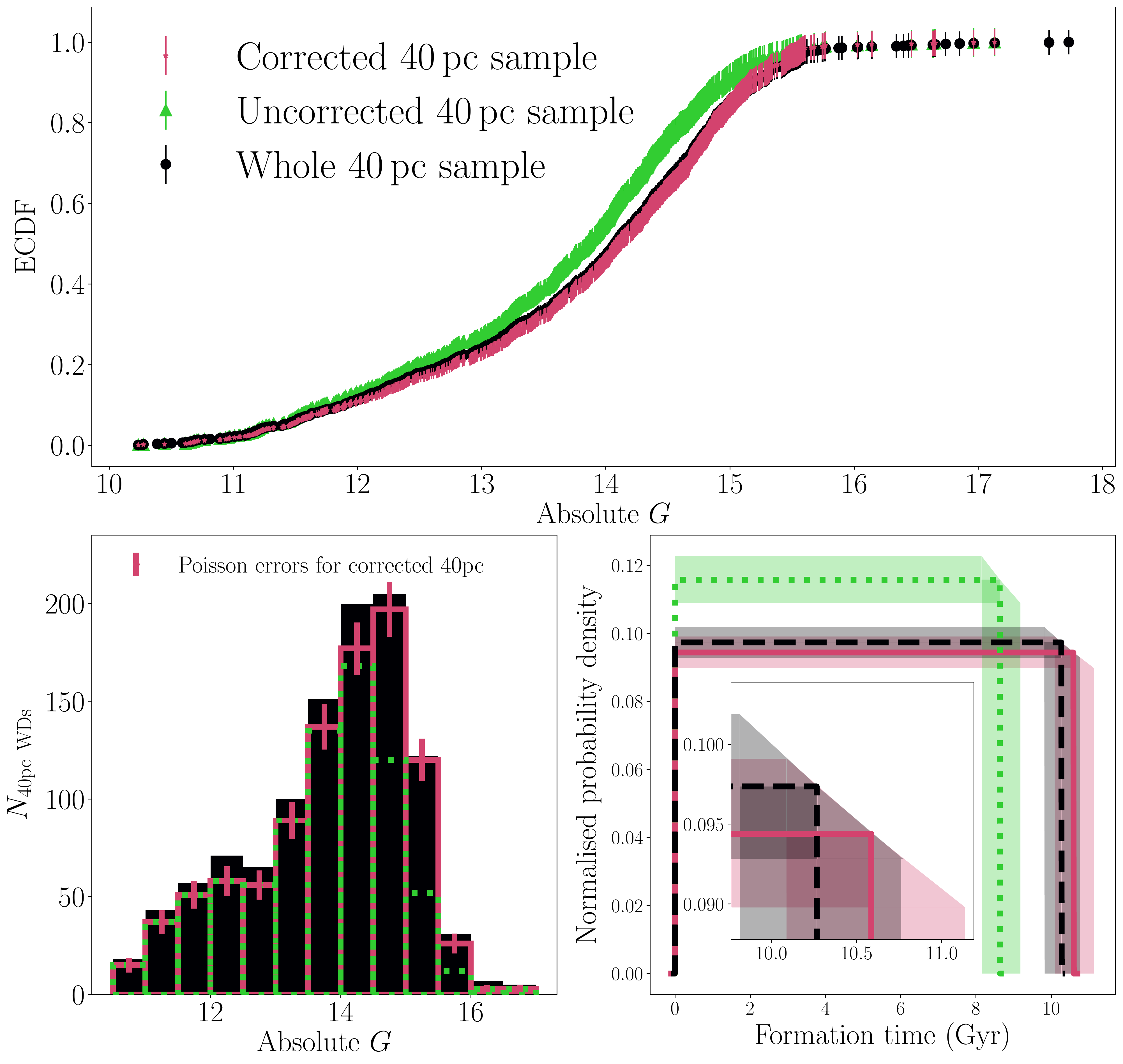}
    \caption{\textit{Top plot}: A comparison of ECDFs for the 40\,pc sample with different mass cuts and corrections applied. The pink points represent the 40\,pc sample we have used in the rest of the paper, in this plot we refer to it as the corrected 40\,pc sample. This sample was corrected using the correction in Table~\ref{tab:mass_corr} and any white dwarf with a mass below the single star formation mass limit was removed. The green triangles represent the sample where the mass corrections were not applied, but any white dwarf with a mass below the single star formation mass limit was removed. The black dots represent the complete 40\,pc sample with no mass correction and no double degenerate removal, even the white dwarfs whose mass is impossible if one assumes single star formation.
    \textit{Bottom left plot}: The histograms of the absolute $G$ magnitude for the samples introduced in the top plot. The same colours are used to represent the different samples as in the top plot.
    \textit{Bottom right plot}: The best fitted uniform, one-component stellar formation histories found from fitting the different variations of the 40\,pc sample. The same colours represent the samples defined in top plot. This plot is similar to the left plots shown in Fig.~\ref{fig:met_imf_taums},~\ref{fig:ifmr_vel_bin_db} and~\ref{fig:argentina}.}
    \label{fig:mass_corr_tests}
\end{figure*}

In the bottom right plot of Fig.~\ref{fig:mass_corr_tests} we show the best fitted uniform, one-component stellar formation histories for all three samples. In all cases the simulation stays the same, only the observed sample is changed when fitting. As before, we used MCMC with 1000 walkers to derive the probability density functions of the best fitted SFHs. Table~\ref{tab:ass} shows the $\chi^2$ and $\tau_{\mathrm{max}}$ of the fits as before. As expected, the "Corrected" and the "Whole" samples result in very similar SFHs. However, the "Uncorrected sample" results in a SFH which starts producing stars closer to present time, since this sample has more bright white dwarfs when compared to the other two samples. In all cases the reduced $\chi^2$ is close to 1, indicating that uniform, one component SFH is a good fit to the data.

Overall, the "Corrected sample" is in better agreement with the population synthesis expectations (i.e. white dwarfs cooling at constant mass) than the "Uncorrected sample" (see Sect.~\ref{sec:40pc} for more detailed discussion). Thus, in the following we proceed with the stellar formation history derived from the "Corrected 40\,pc sample".

\section{Discussion and conclusions}~\label{sec:disc}

We have demonstrated that the white dwarf cooling tables and initial mass function are the main assumptions that could allow for a deviation from uniform stellar formation history with one galactic component. 
We conclude that the white dwarf cooling models of \cite{bedard2020} support a uniform stellar formation history for the 40\,pc sample, but this may not be the case if other white dwarf cooling models such as the LPCODE were to be used. 

By comparing our uniform, one Galactic component SFH with other SFHs we can conclude that some minor $\approx$ 50\% deviation from uniformity cannot be ruled out, such as the SFH of \cite{reid2007} or \cite{mor2019}. However, more peaked SFHs of \cite{tremblay2014} and \cite{fantin2019} are rejected by the data. The studies of \cite{tremblay2014} and \cite{fantin2019} both use the same white dwarf cooling tables as us, albeit older versions. As such, it is unlikely that the disagreement with those SFHs is due to the cooling tables. 

It is important to note that Sect.~\ref{sec:fitting} has shown that increasing the number of Galactic components could also improve the fit without needing to change other astrophysical parameters, and still assuming separate, but uniform SFHs for each component. Therefore, an argument can be made that any small deviation from the constant formation history derived in this work can be compensated by a small change in the input astrophysics. External constraints on these parameters, such as accurate radial velocities for all stars in the sample, better white dwarf cooling models or better constraints on the IMF, will be needed to break this degeneracy. 

Our conclusion is that the 40\,pc sample of white dwarfs supports a uniform stellar formation history given the current error bars on the various astrophysical relations relevant for predicting the properties of the local white dwarf population. Studies in need of a stellar formation history can safely use the assumption of a constant formation history, with the onset of stellar formation (age of the disc) and the number of Galactic components within the local population, being the only parameters which remain difficult to constrain. 

The onset of stellar formation ($\tau_{\rm max}$) in our local region of the Milky Way is linked to over two decades of work on the determination of the age of the Galactic disc from white dwarf luminosity functions. Our main result is that several significant uncertainties remain in the estimation of this parameter. We have shown that changes in the IMF and white dwarf cooling models can all change this parameter by up to $\pm1$~\,Gyr. Furthermore, our tests with different white dwarf cooling models is far from exhaustive. Recently, it has been shown that our theoretical understanding of white dwarf cooling is still lacking in some respects, as a missing cooling process related to crystallisation has been identified when comparing theoretically predicted and observed white dwarf luminosity functions \citep{tremblay2019nat,cheng2019,cheng2020,bauer2020,blouin2021,camisassa2021}. Therefore, we refrain from updating the age of the Galactic disc based on our results, but note that our standard simulation finds a value of 10.6 $\pm$ 0.5\,Gyr, and that additional systematic uncertainties may be as much as $\pm1$~\,Gyr.

\section*{Acknowledgements}
The authors would like to thank the anonymous referee for their helpful comments.
EC and PET were supported by grant ST/T000406/1 from the Science and Technology Facilities Council (STFC). PET has received funding from the European Research Council (ERC) under the European Union’s Horizon 2020 research and innovation programme (Grant agreement No. 101020057). This work has made use of data from the European
Space Agency (ESA) mission Gaia (\url{https://www.cosmos.esa.int/gaia}), processed by the Gaia Data Processing and Analysis
Consortium (DPAC, \url{https://www.cosmos.esa.int/web/gaia/dpac/consortium)}. Funding for the DPAC has been provided by
national institutions, in particular the institutions participating in the
Gaia Multilateral Agreement. 

\section*{Data Availability Statement}
The observational data used in this article are published in \cite{mccleery2020,gentilefusillo2021} and \cite{obrien2022}. The results derived in this article will be shared on a reasonable request to the corresponding
author. 




\bibliographystyle{mnras}
\bibliography{aamnem99,aabib_elena} 






\bsp	
\label{lastpage}
\end{document}